\documentclass[
reprint,
superscriptaddress,
nofootinbib,
amsmath,amssymb,
aps,
prd,
]{revtex4-2}

\usepackage{graphicx}
\usepackage{dcolumn}
\usepackage{bm}
\usepackage{gensymb}
\usepackage{subfigure}
\usepackage{amsmath,siunitx}
\usepackage{hyperref}
\usepackage{booktabs, multirow}

\begin{document}

\preprint{APS/123-QED}

\title{Identifying Black Holes Through Space Telescopes and Deep Learning}


\author{Yeqi Fang}
\email{fangyeqi@stu.scu.edu.cn}
\affiliation{Center for Theoretical Physics, College of Physics, Sichuan University, Chengdu, 610065, China}

\author{Wei Hong}
\email{weihong@mail.bnu.edu.cn}
\affiliation{Department of Astronomy, Beijing Normal University, Beijing 100875, People's Republic of China}

\author{Jun Tao}
\email{taojun@scu.edu.cn}
\affiliation{Center for Theoretical Physics, College of Physics, Sichuan University, Chengdu, 610065, China}



\begin{abstract}
	The EHT has captured a series of images of black holes.
	These images could provide valuable information
	about the gravitational environment
	near the event horizon. However, accurate detection and parameter estimation for candidate black holes are necessary.
	This paper explores the potential for identifying black holes in the ultraviolet band using space telescopes. We establish a data pipeline for generating simulated observations and present an ensemble neural network model for black hole detection and parameter estimation. For detection tasks, the model achieves mean average precision [0.5] values of 0.9176 even when reaching the imaging FWHM ($\theta_c$) and maintains the detection ability until $0.54\theta_c$.
	For parameter estimation tasks, the model can accurately recover the inclination, position angle, accretion disk temperature and black hole mass.
	These results indicate that our methodology can go beyond the limits of the traditional Rayleigh diffraction limit and enable super-resolution recognition.
	Moreover, the model successfully detects the shadow of M87* from background noise and other celestial bodies, and estimates its inclination and position angle.
	Our work demonstrates the feasibility of detecting black holes in the ultraviolet band and provides a new method for black hole detection and further parameter estimation.

\end{abstract}
\maketitle


\section{INTRODUCTION}


In April 2019, the Event Horizon Telescope (EHT) Collaboration released the first shadowed image of M87* \cite{EHT1, EHT2, EHT3, EHT4, EHT5} and in May 2022, they released images of Sagittarius A* \cite{SgrA, SgrA2, SgrA3, SgrA4}, the black hole at the center of the Milky Way. These images provide concrete evidence of the existence of black holes, which is a key tenet of general relativity \cite{Schwarzschild}.


The event horizon of a Schwarzschild black hole is defined by $r_s=GM/c^2$, where $G$ is the gravitational constant, $c$ is the speed of light, and $M$ is the mass of the black hole. Any particle (including photons) that enters this range will inevitably fall into the black hole's singularity. However, that does not mean a black hole can not be observed using a telescope. We can still observe it through its accretion disk, which is the ring of gas and dust surrounding a black hole. Objects falling into the black hole are subjected to the strong gravitational force of the black hole and then rotate around it at high speed while being heated to extremely high temperatures and emitting electromagnetic waves \cite{Accetion1}. The projection of its unstable photon region on an observer's sky is called a black hole shadow \cite{perlick2022calculating}. Accretion disks emit light across many wavelengths. For most black holes in the universe ($\sim10M_{\odot}$), the radiation consists mainly of X-rays, but for larger mass black holes ($\sim10^4M_{\odot}$), the main electromagnetic waves radiated are ultraviolet (UV) to X-rays \cite{2012MNRAS.420..684P}. For supermassive black holes such as M87* and Sagittarius A*, the main mode of radiation is synchrotron radiation, which falls inside the radio band wavelength \cite{yang2020radio}.


The EHT has tested the probability of detecting black holes using a radio interferometer \cite{EHT}. With the development of interferometers, optical interferometer arrays such as COAST \cite{baldwin1996first}, NPOI \cite{armstrong2013navy} and IOTA \cite{carleton1994current} have achieved higher resolution in infrared and even visible wavelengths. However, some smaller black holes might emit higher-frequency waves \cite{Accetion}, which are out of the observable range of radio and optical interferometers \cite{quirrenbach2001optical}. Therefore, these black holes are better observed using optical telescopes, which can cover visible and UV light. Among the candidate wavelengths, the short wavelength of UV light corresponds to higher imaging resolution. Moreover, compared to X-rays and $\gamma$-rays, UV is easier to be focused by optical instruments, making it possible for humans to detect black holes in this band. At present, some UV space telescopes have been successfully launched and operated, such as the Ultra Violet Imaging Telescope (UVIT) \cite{UVIT}, Far Ultraviolet Spectroscopic Explorer (FUSE) \cite{FUSE}, Hubble Space Telescope \cite{Hubble} and so on.


The black hole shadow provides valuable information about the gravitational environment on event horizon scales, enabling verification or modification of general relativity \cite{He_2022, Wen_2023, PhysRevD.104.124063, Meng:2003ry}. High accuracy is crucial for both the detection and parameter estimation of the black hole \cite{Doeleman_2008, 2014ApJ7847B, Johannsen_2016}. According to Torniamenti et al \cite{near}, black holes may exist as close as 80~pc from Earth, within the observational range of optical telescopes.
Some evidence also supports that there are black holes within several hundred pc from Earth, including in binary systems within the solar neighborhood \cite{Chakrabarti_2023}.
However, they may be hidden in a large number of images from current space telescopes.  Distinguishing them from other celestial bodies is challenging due to their far distance and proximity to other objects. Moreover, the diffraction limit presents a fundamental constraint on the resolution of optical telescopes, requiring more accurate detection and recognition methods.
This is where machine learning (ML) can be useful \cite{ML}. Sophisticated ML algorithms enable astronomers to automatically search for celestial objects and enhance the resolution of astronomical images beyond what is possible with conventional optics alone \cite{rs15112853}. Techniques such as super-resolution imaging and image reconstruction algorithms trained on simulated data enable astronomers to effectively enhance the resolution of telescope images, offering a glimpse into previously unseen details of celestial objects \cite{Soo_2023}. ML is a powerful tool for addressing various astronomical physics issues, and neural networks (NNs) are increasingly being used for this purpose. For instance, they have been instrumental in improving the resolution of the M87* image \cite{Medeiros_2023} and is used for the identification and classification of celestial objects such as galaxies, star, and supernovae \cite{Wang2018}. In addition, machine learning methods are aiding in the identification of infrequent and hard-to-find astronomical occurrences by analyzing large datasets to uncover subtle patterns and signals that may otherwise be overlooked. \cite{Carrasco_Kind_2013}.

In recent years, convolutional neural networks (CNNs) have been considered one of the most effective tools in the field of image recognition \cite{CNN1}, and have an increasingly wide range of applications in the field of astrophysics, such as the detection of strong gravitational lenses \cite{CNN2} by deep CNNs, the input of time-domain spectrograms into CNNs for the detection of gravitational waves \cite{Extraction}, the detection and classification of gravitational waves \cite{cls}, gravitational wave noise reduction \cite{denoise} and so on. CNNs have also been used to identify black holes in radio telescope observation images and recover black hole parameters \cite{horizon} such as accretion rate, inclination, and position angle. In Ref. \cite{UV}, telescope observation images are mapped to the U-V plane and then recognized by CNNs.

\begin{figure}[htbp]
	\centering
	\includegraphics[width=0.47\textwidth]{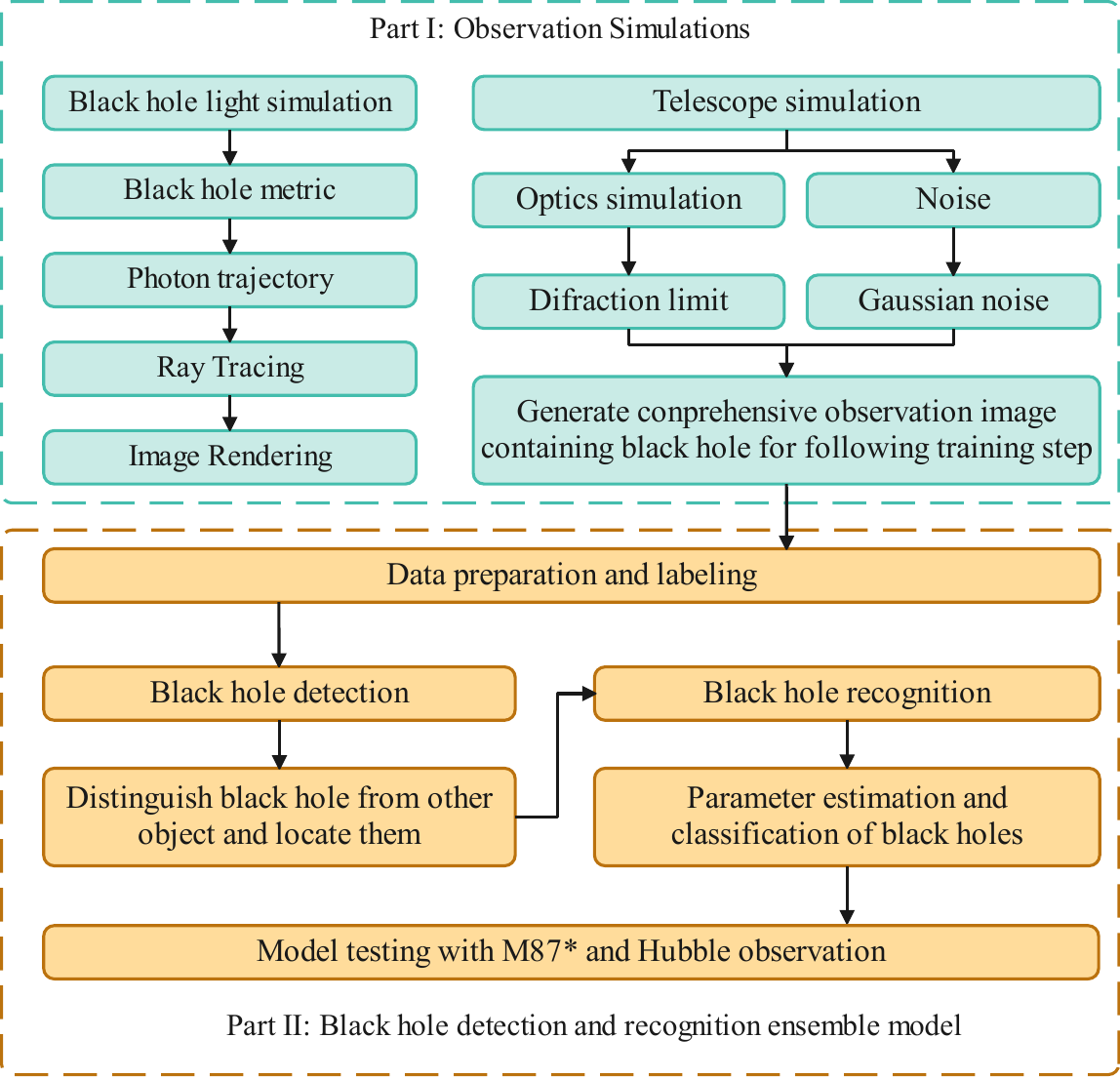}
	\caption{Data simulation pipeline and ensemble NN model.}
	\label{fig:flow}
\end{figure}

For black hole simulations, previous studies for radio band observation often use general {  relativistic magnetohydrodynamics} (GRMHD) to simulate the accretion disk and then generate images of black hole shadows \cite{gammie2003harm}. In the imaging of Sgr A*, the EHT collaboration constructs the relationship between theoretical black hole shadows and the observation of ring-like images using a library of simulations and then uses the CLEAN algorithm and Bayesian method to estimate the parameters as well as the confidence level \cite{moscibrodzka2016general, davelaar2018general}.

Unlike the above methods, what we use in this paper is an ensemble model for both detection and parameter estimation. We first calculate the trajectory of photons in relativistically curved spacetime and then render the image by ray-tracing methods \cite{early, EHT5, shadow} to establish the data pipeline for the subsequent model. Then we present an ensemble NN model with the backend of You Only Look Once (YOLO) \cite{yolov1} and EfficientNet \cite{EfficientNet}. For black hole detection, our detector can accurately distinguish black holes in observation images from tens to hundreds of other celestial objects and determine their positions in the image with a confidence level.
For parameter estimations, it can infer the parameter of black holes from the shadow, where four parameters are selected, including inclination $i$, mass of black hole $M$, position angle $\phi$, and accretion disk temperature $T$.

This paper is organized as follows: In section~\ref{sec:simulation} we render black hole accretion disks using ray tracing and then use the simulated telescope to get the observation images. In section~\ref{sec:model} we introduce the ensemble NN model for both detection and parameter estimation of black holes. In section~\ref{sec:results} we test the validity of our model using the image of M87* {and observation from Hubble space telescope}. Finally, in section~\ref{sec:discussion}, we summarize the results and discuss the feasibility of real-time detecting black holes of candidate black holes and further parameter estimation. The flow chart of the whole work is shown in Fig.~\ref{fig:flow}.

\section{OBSERVATION SIMULATION}\label{sec:simulation}

\subsection{Black hole accretion disk simulation}

\begin{figure}[htbp]
	\centering
	\includegraphics[width=0.47\textwidth]{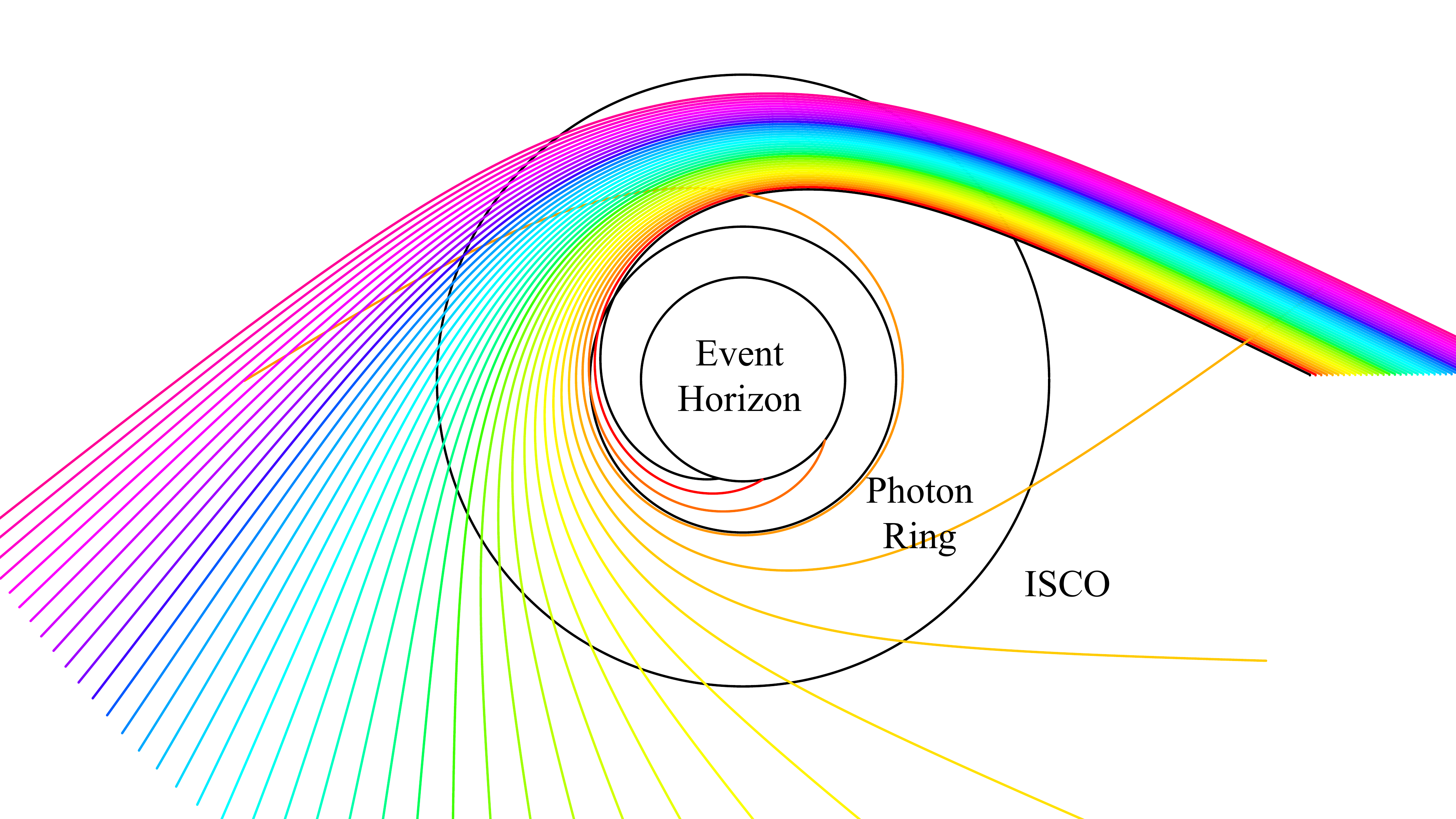}
	\caption{Photon trajectories, event horizont ($r_s$), photon ring ($1.5r_s$) and ISCO, where the initial direction makes 26.5\degree angle with horizontal axis. Note: the colors are only for demonstration and don't indicate the wavelengths.}
	\label{fig:photon}
\end{figure}

To render the image of black holes, ray-tracing for the accretion disk of the Schwarzschild black hole is used, whose metric has the form,
\begin{equation}
	ds^2=-\left(1-\frac{r_{\mathrm{s}}}{r}\right) c^2 d t^2+\left(1-\frac{r_{\mathrm{s}}}{r}\right)^{-1} d r^2+r^2 d \Omega^2,
	\label{eq:metric}
\end{equation}
where $r_s=GM/c^2$ and $d \Omega^2=\left(d \theta^2+\sin ^2 \theta d \phi^2\right)$. From this equation, the photon trajectories outside the black hole can be solved numerically using the fourth-order {  Runge-Kutta} algorithm with $\theta=\pi / 2$,
\begin{equation}
	\frac{d^2u(\phi)}{d\phi^2}=\frac{3}{2} u(\phi)^2-u(\phi),
	\label{eq:de}
\end{equation}
where $u(\phi)=1/r(\phi)$. The result is shown in Fig.~\ref{fig:photon}. The innermost stable circular orbit (ISCO) is the smallest edgewise-stable circular orbit in which particles can be stabilized to orbit a massive object in the theory of general relativity. No particle can maintain a stable circular orbit smaller than $r_{\text{ISCO}}$. In that case, it would fall into the event horizon of the black hole while rotating around it. For a Schwarzschild black hole, $r_{\text{ISCO}}=3r_s$. Typically, this is where matter can generate an accretion disk \cite{Kawashima_2019, Dokuchaev_2019, Krolik_2002}, which corresponds approximately to the center of the accretion disk in this work.

The temperature of the accretion disk determines the wavelength of black body radiation, which in turn determines whether a black hole can be observed through a telescope within a certain wavelength range. The temperature of the accretion disk is \cite{2012MNRAS.420..684P, Accetion}:
\begin{equation}
	T=\left[6.27 \times 10^7 \mathrm{~K}\right]\alpha^{\frac{1}{4}} \left(\frac{M}{3M_{\odot}}\right)^{-\frac{3}{4}} \left(\frac{\dot{M}}{10^{17}\mathrm{g}/\mathrm{s}}\right)^{\frac{1}{2}},
	\label{eq:temp}
\end{equation}
where $M_{\odot}$ is the {  solar mass}, $\dot{M}$ is the accretion rate and $\alpha$ is the standard alpha viscosity, and $0 < \alpha < 1$ is a dimensionless coefficient, assumed by Shakura and Sunyaev to be a constant \cite{1973AA337S}. To reduce dimensions of the parameter space, we set $\alpha=0.1$ and $\dot{M}=7.26\times10^{16}g$ $\text{sec}^{-1}$.
We can assume that the accretion disk is radiatively efficient, i.e. the rate of accretion is small enough so that any heat generated by viscosity can be immediately converted into light energy and radiated outward. It is also supposed that the accretion disk is very thin, resulting in all accreted material being on the equatorial plane. \cite{Accetion}.

To render a more realistic image of the black hole, gravitational lensing \cite{cunha2018shadows} and the Doppler effect should also be considered \cite{james2015gravitational}. The Doppler color shift is given by
\begin{equation}
	(1+z)_{\text{Doppler}} =\frac{1-\beta \cos(\gamma)}{\sqrt{1-\beta^2 }},
	\label{eq:doppler}
\end{equation}
where $\gamma$ is the angle between the ray direction and the disk's local velocity \cite{shadow}.
The redshift from the relative motion of the black hole to the observer can be ignored since in our simulations the Earth is typically about several hundred light-years away from the black hole.

\begin{figure*}[htbp]
	\centering
	\includegraphics[width=0.95\textwidth]{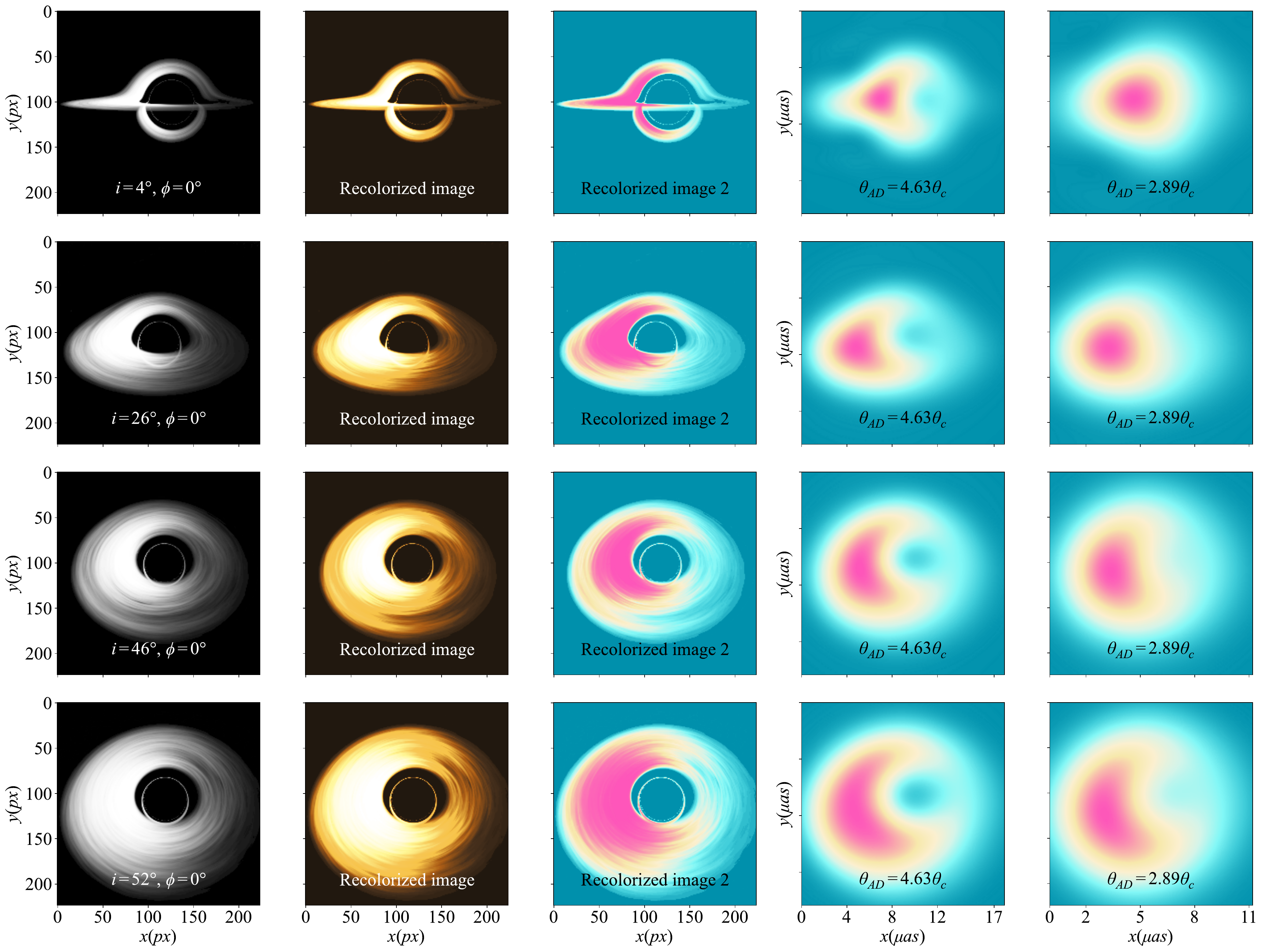}
	\vspace{-0.3cm}
	\caption{Simulated image, wherein the first column is the simulated generated image, the second and third column is the recolored image, and the fourth and fifth columns are the blurred images in the simulated telescope.}
	\label{fig:black_boles}
\end{figure*}

For simplicity, one can consider blackbody radiation and disregard other radiation, such as synchrotron radiation. According to Planck's formula for blackbody radiation \cite{Bramson1968}, it can be calculated that the intensity of radiation at a certain wavelength is $f(\lambda) = \frac{1}{\lambda^5} \frac{1}{\exp(hc/\lambda k_B T-1)}$. Since we assume that the telescope operates at a single wavelength, the brightness observed by the telescope is also proportional to $f(\lambda)$, and to simplify the calculations, we ended up simplifying the telescope photo to a black-and-white photo and normalizing the radiant intensity over [0, 255]. The result is shown in the first column of Fig.~\ref{fig:black_boles}.
Note that the radiation used to simulate the black body of a black hole is UV light with a narrow spread of wavelengths. Therefore it can be seen as monochromatic light, so the image is shown in grayscale. To demonstrate intuitively, it is mapped to be a colored image in the second and third column of Fig.~\ref{fig:black_boles}. We can see that the light that black holes emit is not symmetrical. That is because the gravitational force of the black hole bends the light, which makes the accretion disk twist into the shape of a ``mushroom''.

We can simulate the star through the star mass-luminosity relation \cite{Salaris}:
\begin{equation}
	\frac{L}{L_{\odot}} =\left(\frac{M}{M_{\odot}}\right)^a, \quad 0.43 M_{\odot}<M<2 M_{\odot},
\end{equation}
where $M_{\odot}$ and $L_{\odot}$ are the mass and luminosity of the Sun and $1<a<6$. We take $a=4$ in this simulation, which is the most probable range for star in the universe.



\subsection{Telescope simulation}
The diffraction limit is the fundamental constraint on telescope resolution. According to the Rayleigh criterion, two objects are considered just resolvable if the first minimum (dark fringe) of the diffraction pattern created by one object coincides with the central peak of the pattern created by the other. The imaging FWHM of a telescope is $\theta_c=\frac{1.22 \lambda}{\mathrm{D}}$, where $\lambda$ is the wavelength and D is the diameter of the telescope. Throughout the observing range of optical telescopes, UV light has the highest resolution. Electromagnetic waves with smaller wavelengths, such as X-rays and $\gamma$-rays, are no longer possible for an optics telescope to observe because of the difficulty in focusing. To prevent atmosphere absorption of UV light, a telescope has to be placed on satellites. In our work, the configuration of the simulated telescope follows the Hubble Telescope \cite{Hubble_web}, with the imaging FWHM of 10 $\mu$as (1000$\mu$as = $\SI{1}{\arcsecond}$), as shown in Table~\ref{tab:tele_para}.

\begin{table}[htbp]
	\centering
	\caption{Telescope configuration}
	\begin{tabular}{l@{\hskip 0.1in}@{\hskip 0.1in}l@{\hskip 0.2in}l}
		\hline \hline \addlinespace[2pt]
		Symbol            & Value                 & Explanation                        \\
		\addlinespace[1pt]\hline\addlinespace[2pt]
		D                 & 2.4m                  & Diameter                           \\
		F                 & 57.6m                 & Focal length                       \\
		$L_{\text{CCD}}$  & $2\mu m$              & Size of the pixels on the detector \\
		$ N_{\text{CCD}}$ & 3072                  & Number of pixels of the CCD        \\
		SNR               & 10                    & Signal-to-Noise Ratio              \\
		$\theta_c$        & \SI{0.01}{\arcsecond} & Angular resolution in arcsecond    \\
		\hline \hline
	\end{tabular}
	\label{tab:tele_para}
\end{table}

\begin{figure}[htbp]
	\centering
	\includegraphics[width=0.47\textwidth]{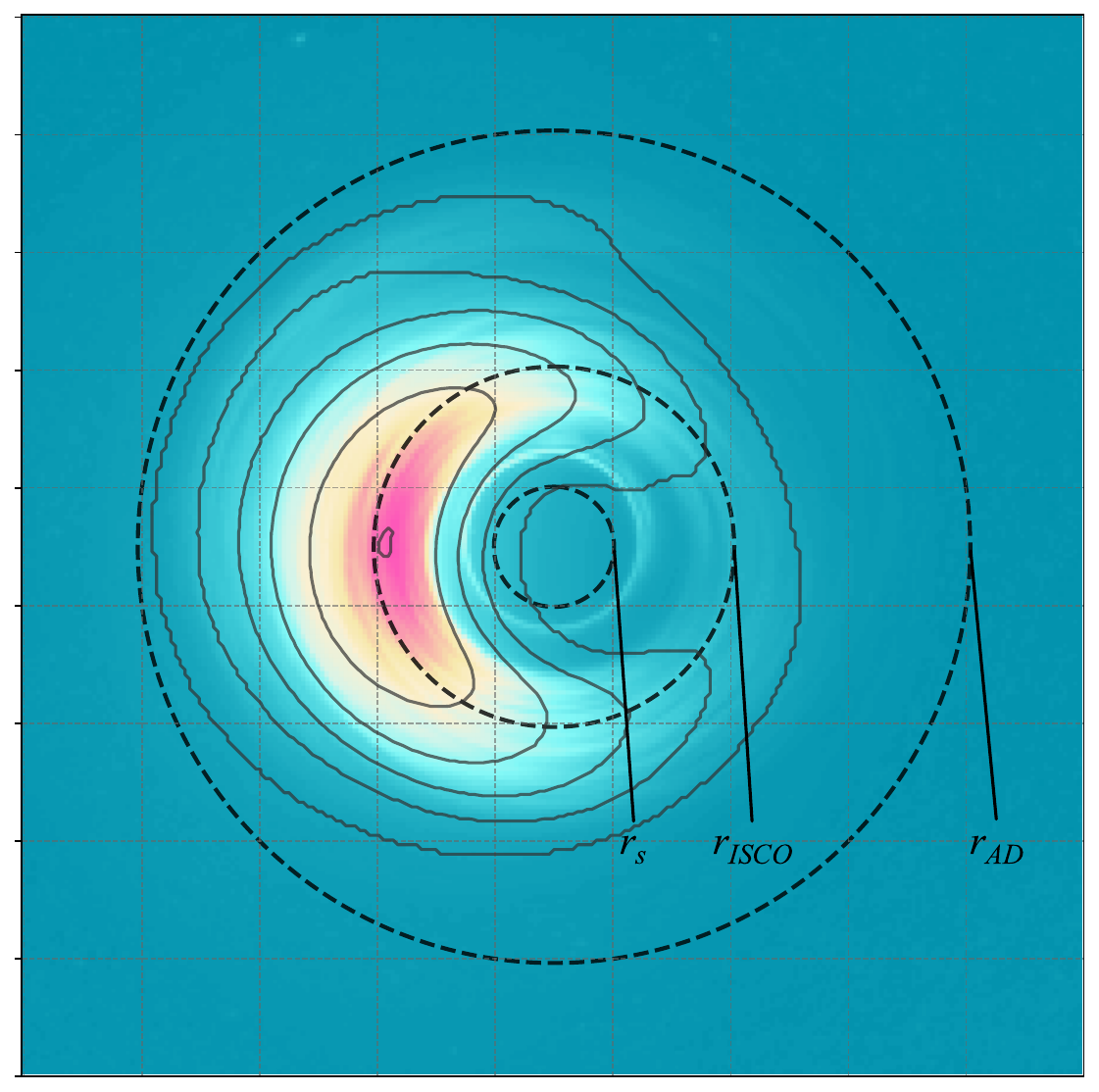}
	\caption{Schematic of the size relationship between the black hole event horizon $r_s$, radius of ISCO $r_\text{ISCO}$ and radius  of accretion disk $r_\text{AD}$, where $\theta_\text{AD}=14.8\theta_c$, $i=32.4\degree$ and $\phi=0\degree$.}
	\label{fig:diff2}
\end{figure}

After generating the simulated image of the black hole and star, the Point Spread Function (PSF) of the telescope for different angular sizes of images is calculated.
{The PSF describes the response of our telescope to a point source or point object. It essentially characterizes how a point light source would appear in the image, taking into account the diffraction effects, aberrations, and other imperfections of the optical system. In our situation, only diffraction is considered.}
Then, the PSF of the telescope is convolved with the simulated image to obtain the observed results. This process is shown in Fig.~\ref{fig:telescope} (a)-(c). The shadows with different angular sizes are shown in Fig.~\ref{fig:telescope} (d)-(h). We define the angular size of the {  input image of the model} as $\theta$, the angular size of the outer edge of the accretion disk as $\theta_\text{AD} = \arcsin\left(2 r_\text{AD} / r_\text{obs}\right)$, and the angular size of ISCO as $\theta_\text{ISCO}=\arcsin{(2 r_\text{ISCO} / r_\text{obs})}$, where $r_\text{obs}$ is the distance between the black hole and the observer.
The doughnut-like shadow and size relations are shown in Fig.~\ref{fig:diff2}. There is almost no light distribution inside the event horizon ($r_s$). The ISCO ($3r_s$) is approximately the center of the accretion disk and $r_\text{AD} \approx 2 r_\text{ISCO}$.

When $\theta_{\text{ISCO}}>\theta_c$, the shadow is a doughnut-shaped bright spot with unequal brightness on both sides, which is easy to distinguish, as shown in Fig.~\ref{fig:telescope} (e)(f)(g). When $\theta_{\text{ISCO}}<\theta_c$, the shadow is connected to form a circular facula, which is difficult to recognize by the naked eye, as shown in Fig.~\ref{fig:telescope} (h).

To match the real observation as closely as possible, noise should also be considered, which is determined by the SNR of the telescope with ${\rm SNR} = N/\Delta N$, where N is the number of photons released by the source, and $\Delta N$ is the noise. In optics and telescopes, the Charge Coupled Device (CCD) serves as a sensitive detector capturing light from celestial objects and converting it into digital signals for analysis. Suppose the number of photo-electrons detected from the object, sky background and dark current is $S_o, S_b$ and $S_d$ respectively, with the time-independent readout noise $R$,
the CCD SNR equation is written as \cite{CCD_SNR}
\begin{equation}
	\mathrm{SNR}=\frac{S_o Q t}{\sqrt{S_o Q t+S_b Q t n_p+S_d t n_p+R^2 n_p}},
\end{equation}
where
$n_p$ is the number of pixels that the object is spread over, $t$ is the exposure time in seconds and $Q$ is the quantum efficiency of the CCD, expressed as a number between 0 and 1.
Referring to the parameters of the Hubble Telescope as well as its historical observations \cite{Hubble_web}, we use Gaussian noise and make all simulated observations satisfy $\text{SNR}< 10$.







\section{ENSEMBLE MODEL FOR DETECTION AND RECOGNITION}\label{sec:model}
To ensure clarity and coherence, it is essential to introduce some concepts relevant to our discussion. \textbf{Detection} refers to the model's ability to identify black holes in observation images. This includes distinguishing black holes from other celestial objects and locating their positions.
\textbf{Recognition} involves estimating parameters for both continuous and discrete variables. \textbf{Regression} focuses on predicting continuous variables, while \textbf{classification} focuses on discrete variables.

\subsection{Datasets}

In this paper, two NN models for black hole detection and parameter estimation share the same data generation pipeline but with different configurations. The former corresponds to datasets where black holes and star are generated in one image with the size of $1024\times1024$, while the latter has datasets that fix black holes in the center, with different sizes of accretion disk, inclinations, position angles and temperatures, with an image size of $240 \times240$.

For the detection task, multiple data groups are generated with different $\theta$, each containing 1~000 observation images. Each image has a corresponding text file with metadata on the bounding circles that define the objects in the image. The metadata for each object includes its class, x-y coordinates, and radius of the bounding circle. There is either zero or one black hole and 3 to 100 star in one image.
\begin{figure}[htbp]
	\centering
	\includegraphics[width=0.48\textwidth]{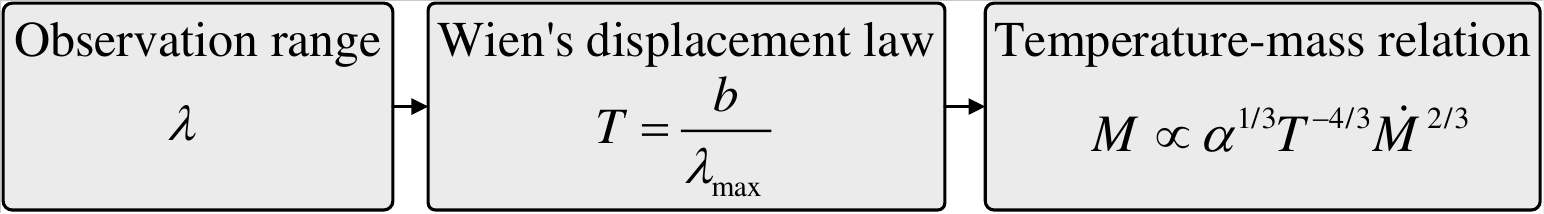}
	\caption{Determining the range of $T$ and $M$ for NN model}
	\label{fig:parameter_range}
\end{figure}

\begin{figure*}[htbp]
	\centering
	\subfigure[Image before convolution]{\includegraphics[width=0.32\textwidth]{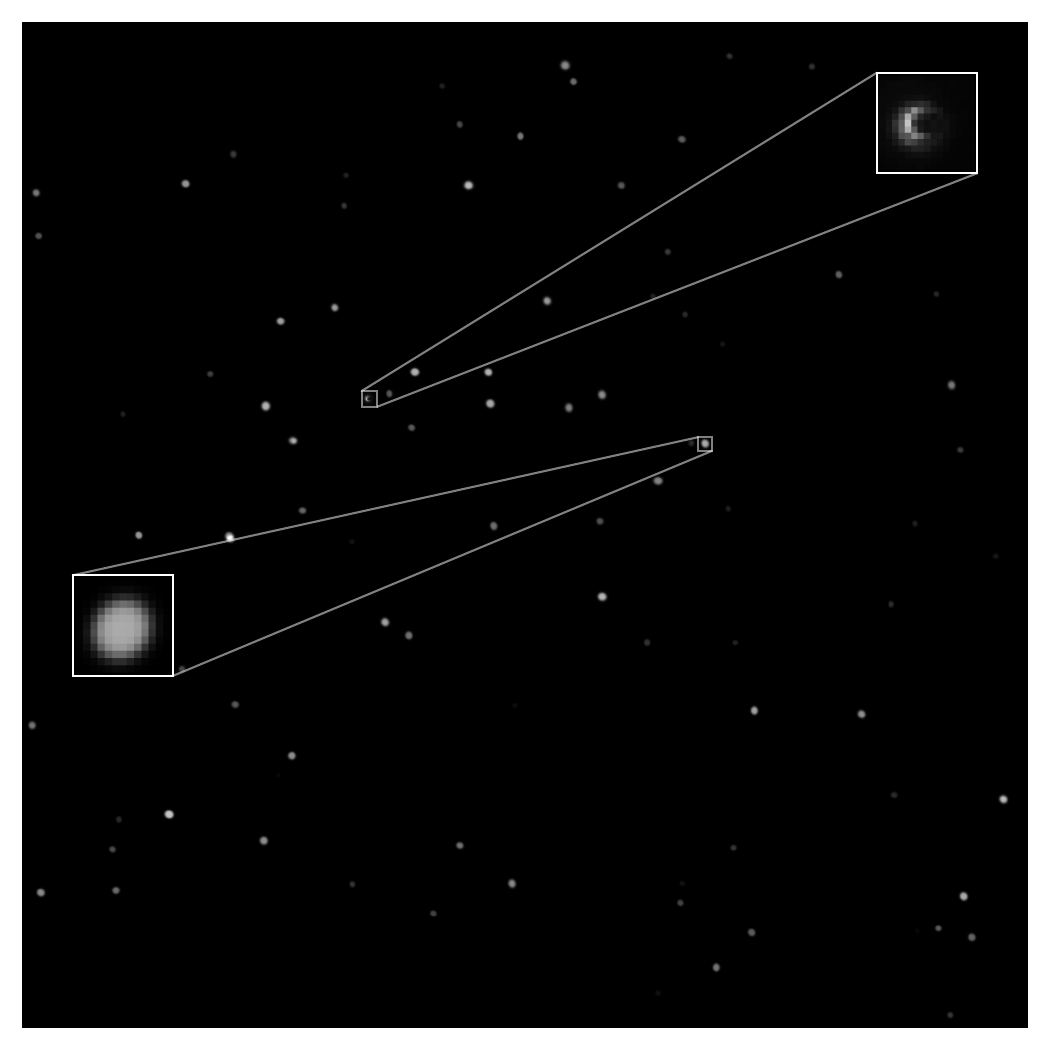}}
	\subfigure[PSF of the telescope]{\includegraphics[width=0.32\textwidth]{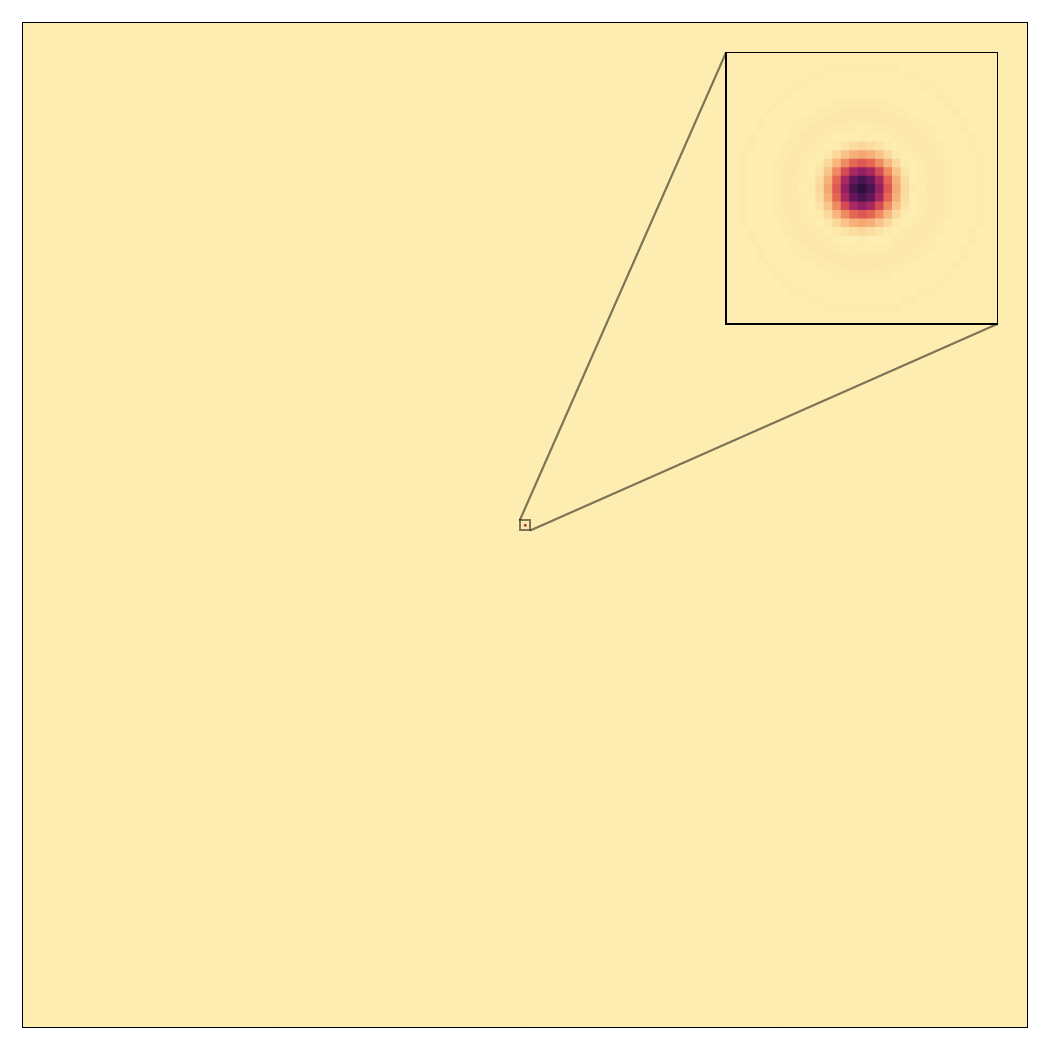}}
	\subfigure[Image after convolution]{\includegraphics[width=0.32\textwidth]{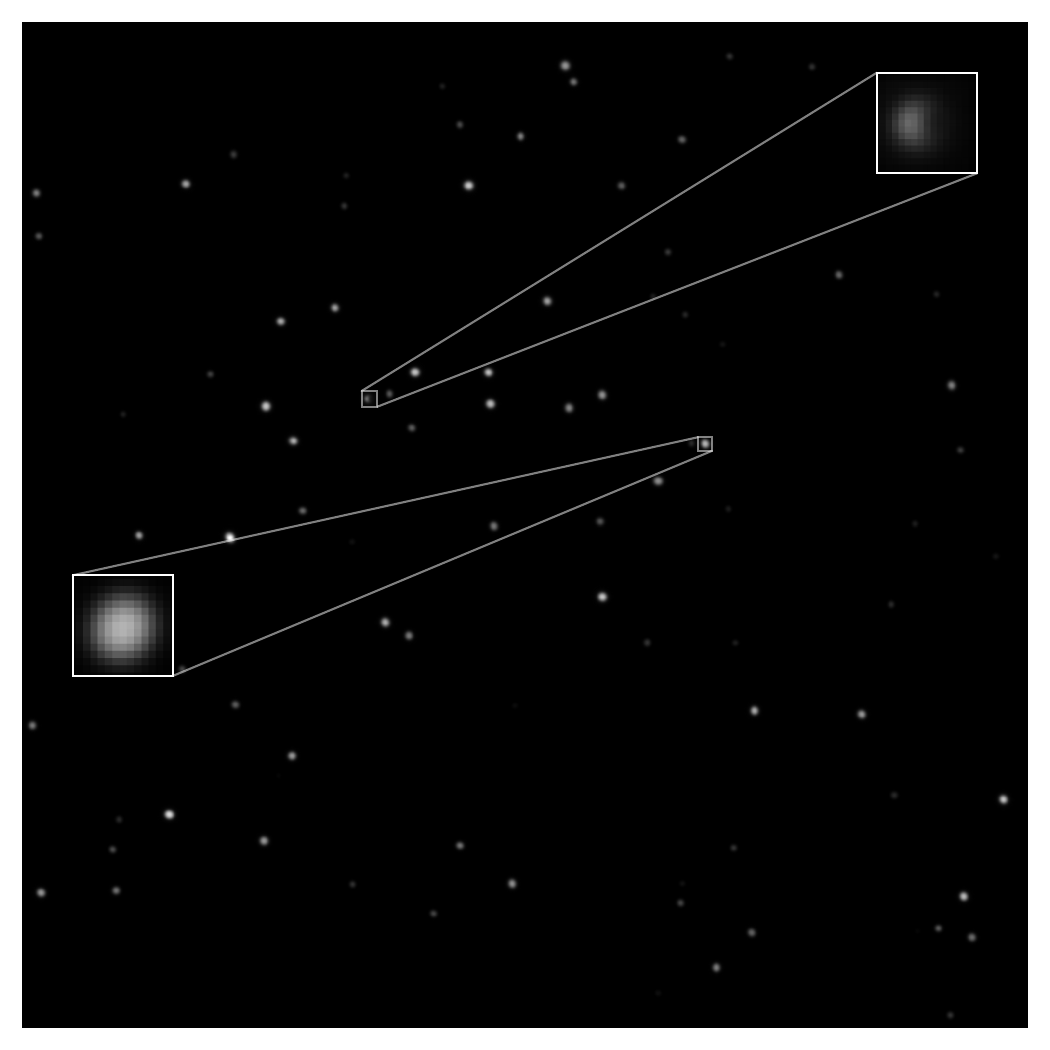}}
	\subfigure[No diffraction]{\includegraphics[width=0.19\textwidth]{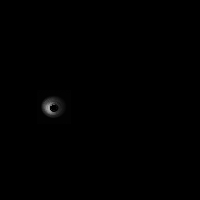}}
	\subfigure[$\theta_\text{AD}=10.1\theta_c$]{\includegraphics[width=0.19\textwidth]{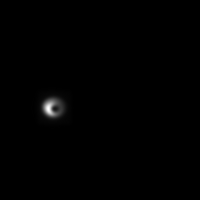}}
	\subfigure[$\theta_\text{AD}=5.4\theta_c$]{\includegraphics[width=0.19\textwidth]{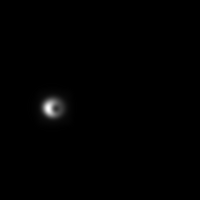}}
	\subfigure[$\theta_\text{AD}=3.1\theta_c$]{\includegraphics[width=0.19\textwidth]{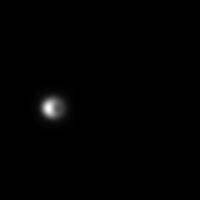}}
	\subfigure[$\theta_\text{AD}=1.2\theta_c$]{\includegraphics[width=0.19\textwidth]{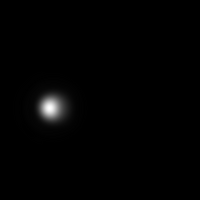}}
	\vspace{-0.3cm}
	\caption{Telescope simulation, where (a) is the image before convolution, (b) is the {point spread function (PSF)} of the telescope and (c) is the simulated observation image generated by convolution of (a) and (b). The image is zoomed in to show the details. In (a) and (c), the left zoomed image is a black hole and the right zoomed image is a star. (d)-(h) is black holes with different angular sizes.}
	\label{fig:telescope}
\end{figure*}

For the parameter estimation task, we also generated several groups of data according to the different $\theta$. Each group has 27~018 images{}. The temperature is determined by Wien's displacement law from the observable range of the telescope, and the mass of a black hole is inferred from its size of accretion disk by Eq.~(\ref{eq:temp}). This process is shown in Fig.~\ref{fig:parameter_range}. The parameters to be estimated and their range are shown in Table~\ref{tab:Classification}.

\begin{table}[htbp]
	\centering
	\caption{Parameter range of the black hole}
	\begin{tabular}{c@{\hskip 0.1in}l@{\hskip 0.1in}l}
		\hline \hline \addlinespace[2pt]
		Parameter & Range                                          & Explanation    \\
		\hline\addlinespace[2pt]
		$i$       & $[-90\degree,90\degree]$                       & Inclination    \\
		$\phi$    & $[0\degree,360\degree]$                        & Position angle \\
		$M$       & $[1\times10^4M_{\odot}, 5\times10^4M_{\odot}]$ & Mass           \\
		$T$       & [1.91 2.06 2.69 3.47] ($\times10^4\mathrm{K}$) & Temperature    \\
		\hline \hline
	\end{tabular}
	\label{tab:Classification}
\end{table}

The generated samples were randomly split into training and validation sets with ratios $N_{\text{train}}: N_{\text{validation}}: N_{\text{test}} = 6:2:2$. To ensure the accuracy of training, the data in the training set is rounded up to an integer multiple of the batch size, and the excess is divided into the validation set.

\subsection{Model introduction}

In computer vision (CV), object detection is typically defined as the process of locating and determining whether specific instances of a real-world object class are present in an image or video.
In recent years, a large range of sophisticated and varied CNN models and approaches have been developed. As a result, object detection has already been widely used in a variety of automatic detection tasks, such as the auto-count of the traffic flow or the parking lot \cite{sultana2020review, zaidi2022survey, dhillon2020convolutional}, making it the best choice for us to detect black holes from the images of telescopes. Among all object detection models, YOLO is considered one of the most outstanding due to its highly accurate detection, classification, and super-fast computation \cite{yolov1}. The YOLO family comprises a series of convolution-based object detection models that have demonstrated strong detection performance while being incredibly light \cite{amit2021object, jiang2022review}. This enables real-time detection tasks on devices with limited computational resources. In particular, we make use of the Ultralytics package for the YOLO model \cite{Jocher_Ultralytics_YOLO_2023}, which implements these models using the Python environment and PyTorch framework. Aside from offering a variety of model architectures with differing pre-trained parameters and sizes of the model, Ultralytics can also provide a wealth of functionality for training, testing, and profiling these models. Various tools are also available for transforming the trained models into different architectures. This facilitates the redesign of our model for the detection of black holes with the YOLO backend.

After obtaining the location of the black hole by the above BH detection model, it is also important to determine the parameters of the black hole and its accretion disk (e.g., $i, \phi, M, T$, etc.), which is also performed by the deep CNN model in this work.
There are lots of famous deep CNNs for image recognition, such as VGG \cite{VGG}, ResNet \cite{ResNet}, DenseNet \cite{DenseNet} and EfficientNet \cite{tan2020efficientnet}. After trial and error for almost all the commonly used CNN models, EfficientNet-b1 turns out to have the highest accuracy and low computational resource consumption.
Similar to YOLO, EfficientNet is a family of models consisting of 8 models ranging from b0 to b7. Each successive model number has more parameters associated with it. In addition to higher accuracy, this model also has a significant advantage in terms of scalability. It is based on the concept of compound scaling, which balances the depth, width, and resolution of the network. This results in a more accurate and efficient model compared to its predecessors. To attain the best outcomes, the model can be scaled by modifying the parameters of EfficientNet to suit the input image's size. This is dissimilar to traditional models that necessitate a uniform input size and may lose information when compressing larger images. However, in astronomical observations, every piece of information is exceedingly valuable and scarce. Therefore, the advent of EfficientNet is a noteworthy advancement. The ideal size of the input image varies from 224 to 600 pixels, from b0 to b7.

\begin{figure*}[htbp]
	\centering
	\includegraphics[width=0.9\textwidth]{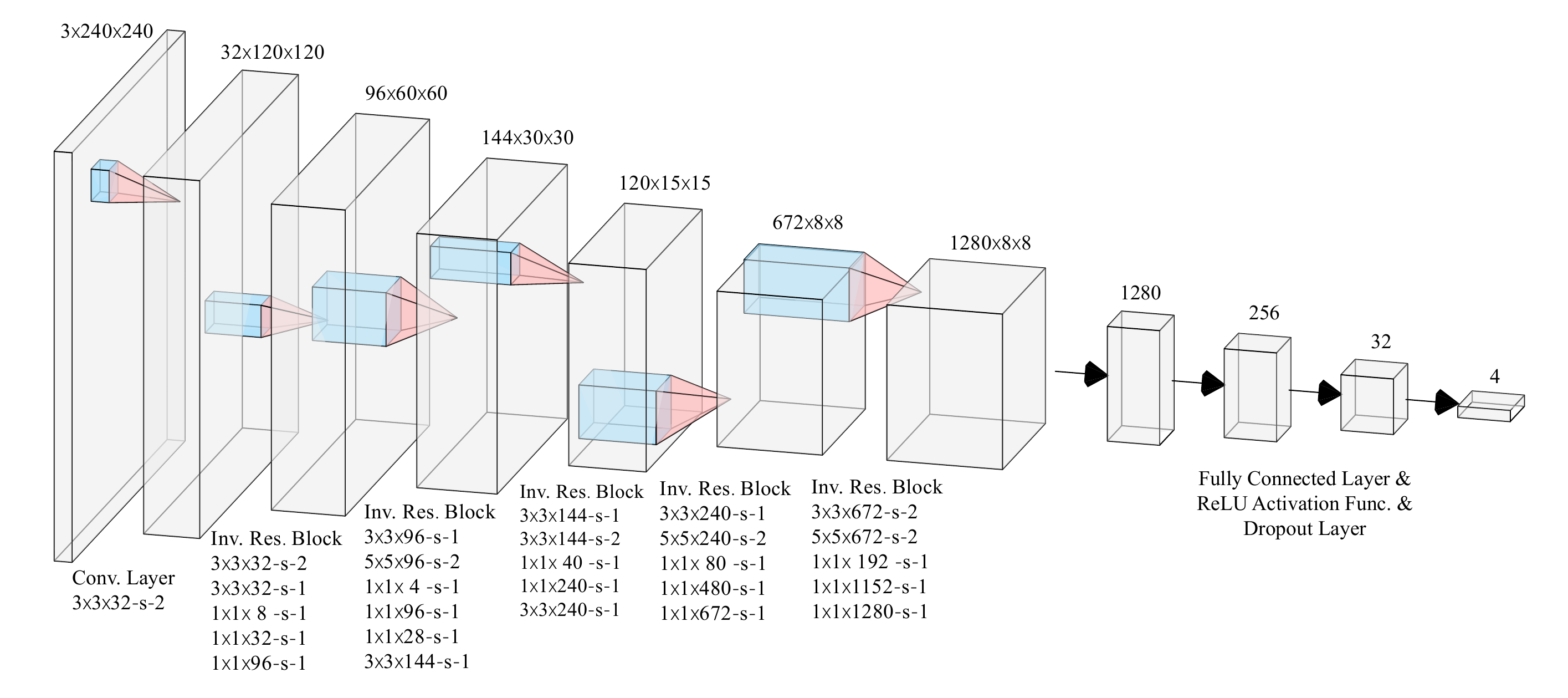}
	\caption{The structure of EfficientNet used in this paper, where the gray tensor denotes the changes of the input image, the blue tensor denotes the convolution kernel, the red arrows denote the convolution process, and the black arrows denote the fully connected layer, with the depth of each tensor and the size of the convolution kernel not plotted to actual scale.}
	\label{fig:nn_structure}
\end{figure*}

\subsection{Method}

\subsubsection{Black hole detection model}\label{sec:detection}

The YOLO v5, v7 and v8 \cite{yolov5, yolov7, yolov8} were trained and tested on the simulated datasets, and YOLOv5 has the best performance in terms of mAP, F$_1$ score and speed.
While the YOLO model is a popular tool for object detection, its application in astrophysics is limited. For example, it uses bounding boxes to locate objects, which is incompatible with circle-based celestial bodies. To address this gap, we have made enhancements to the YOLO backend and developed a specialized model for  detecting circle-shaped celestial bodies for astronomical applications. The computational resources are conserved and accuracy is enhanced by reducing the parameter space to three dimensions (x, y, and radius) compared to the traditional bounding boxes' four dimensions (x, y, width and height). Furthermore, the inherent rotational symmetry of circles ensures consistent results regardless of orientation changes, which is a critical aspect of astronomical observations, such as luminosity calculations. Additionally, the channels of the convolutional kernel have been reduced to handle monochrome imagery, alleviating computational stress. The loss function is changed to circular Intersection over Union (IoU) calculation instead of rectangular to align with the model's focus on circle detection, as explained in the Appendix.

\begin{table*}[htbp]
	\centering
	\caption{{Descriptions and applications of the metrics used in this study.}}
	\begin{tabular}{lp{10cm}@{\hskip 0.1in}p{3.6cm}}
	\hline\hline \addlinespace[2pt]
	Metric & Description & Application in this study \\
	\hline \addlinespace[2pt]
	Precision & The ratio of true positive detections to the total number of positive detections (true positives + false positives). It measures the accuracy of the positive predictions. & Detection \\
	Recall & The ratio of true positive detections to the total number of actual positive instances (true positives + false negatives). It measures the ability to find all relevant instances. & Detection \\
	Accuracy & The ratio of correctly predicted instances (both positive and negative) to the total number of instances. It provides an overall measure of the model's performance. Applied when dataset is balanced. & Detection and classification of $T$ \\
	F1 Score & The harmonic mean of Precision and Recall, providing a balance between the two metrics. It is useful when both Precision and Recall are important, especially for unbalanced datasets. & Detection \\
	mAP$_{[0.5]}$ & Mean Average Precision at IoU threshold 0.5. It evaluates performance of a detection model. & Detection \\
	mAP$_{[0.5:0.95]}$ & Mean Average Precision averaged over multiple IoU thresholds from 0.5 to 0.95. More omprehensive than mAP$_{[0.5]}$ & Detection \\
	MAE & Mean Absolute Error, which measures the average magnitude of errors between predicted and true values. It is used for continuous parameter estimation. & Parameter estimation of $i$, $\phi$, and $M$ \\
	\hline\hline
	\end{tabular}
	\label{tab:metric_description}
\end{table*}

The metrics used in this paper are {listed in Table.~\ref{tab:metric_description}, and the details are} as follows: \textbf{Precision} is calculated as the ratio of true positives (TP, instances correctly identified as positive) to the sum of TP and false positives (FP, instances incorrectly identified as positive).
\textbf{Recall} is calculated as the ratio of TP to the sum of TP and false negatives (FN, instances incorrectly identified as negative),
\begin{equation}
	\text{Precision} = \frac{\text{TP}}{\text{TP + FP}},\quad \text{Recall} = \frac{\text{TP}}{\text{TP+FN}}.
	\label{eq:recall}
\end{equation}

\textbf{Accuracy} is calculated by dividing the total number of instances by the ratio of properly predicted instances. It is the most commonly used metric in classification. However, it may not be suitable for our situations, because there is an imbalanced class distribution. stars are far more than black holes, making accuracy a misleading metric. In contrast, \textbf{F$_1$ score} is suitable to deal with this situation. It is the harmonic mean of precision and recall [cf. Eq.~\eqref{eq:f1}]. It provides a more impartial assessment of the model's efficacy by taking into account both FP and FN,
\begin{equation}
	\text{F}_1=\frac{2 \times \mathrm{Precision}\times \text{Recall}}{\text{Precision}+\text{Recall}}.
	\label{eq:f1}
\end{equation}

\textbf{Intersection over Union (IoU)} is a measure of the overlap between the predicted bounding circle and the ground truth bounding circle. When the IoU is 0.5 or greater, the prediction is considered a true positive. For the detailed formula see Eq.~\eqref{eq:intersection} and \eqref{eq:iou} in the Appendix.

\textbf{Mean Average Precision (mAP)}: There are two versions of mAP, The first one, mAP$_{[0.5]}$, is calculated by considering predictions with an IoU threshold of 0.5 or higher as correct detections.
The mAP$_{[0.5]}$ evaluates how well the algorithm performs when the bounding circles have at least a 50\% overlap with the ground truth. Another version: mAP$_{[0.5:0.95]}$, considers a range of IoU thresholds, specifically from 0.5 to 0.95 with some interval (here we use 0.05 intervals). It provides a more detailed evaluation by taking into account detections at various levels of overlap with the ground truth. So it gives a more comprehensive view of the algorithm's performance across different levels of precision and recall. Considering mAP$_{[0.5:0.95]}$ is more accurate and comprehensive \cite{mAP}, the model is evaluated by 90\% of mAP$_{[0.5:0.95]}$ and 10\% of mAP$_{[0.5]}$.

\begin{figure}[htbp]
	\centering
	\includegraphics[width=0.48\textwidth]{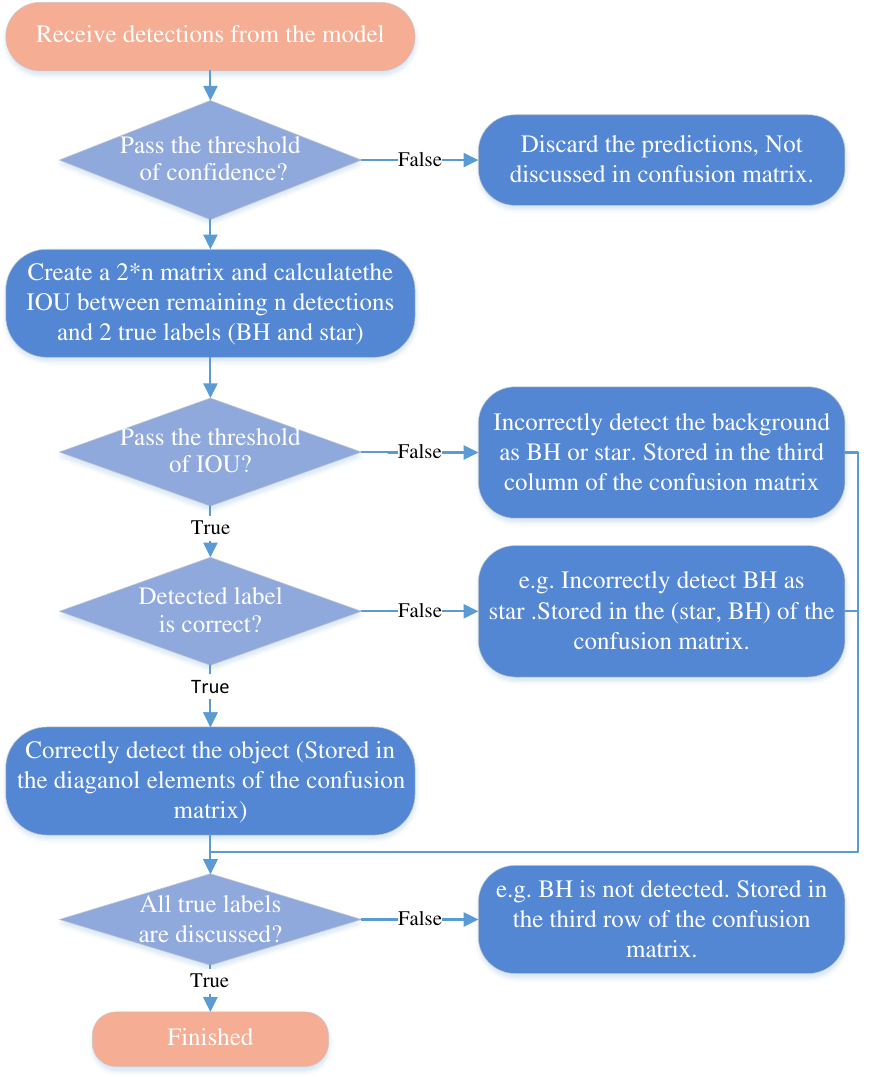}
	\caption{{The flowchart of the black hole detection model. The normalized confusion matrix for the classification of BH, star and background is shown in Fig.~\ref{fig:confusion_matrix}.}}
	\label{fig:flowchart}
\end{figure}

\begin{figure}
	\centering
	\includegraphics[width=0.48\textwidth]{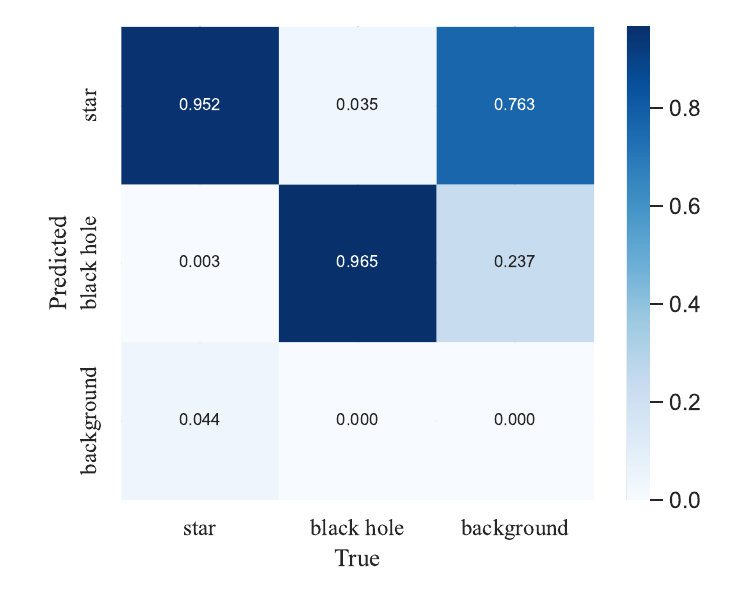}
	\caption{Normalized confusion matrix for the classification of BH, star and background, where the x-axis is the ground truth and the y-axis is the prediction, $\theta_\text{AD}=2.09\theta_c$ and the BH/star ratio is 1/10. For example, the (1, 1) element means that 95\% of the stars are correctly predicted as star, while the (1, 2) element means that 3.5\% of the black holes are incorrectly predicted as star. Each entry is normalized by the sum of the column. Because the model didn't count background predicted as background, only the background misclassified as black hole and star is shown, so the (3, 3) element is labeled as 0\%, which means that among the misclassified background, 76\% are predicted as black hole and 24\% are predicted as star.}
	\label{fig:confusion_matrix}
\end{figure}

\begin{figure*}[htbp]
	\centering
	\subfigure[F$_1$ Score]{\includegraphics[width=0.24\textwidth]{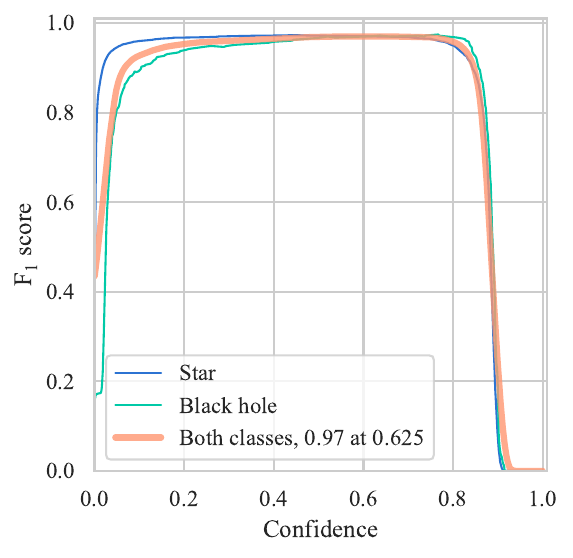}}
	\subfigure[Precision-Recall Curve]{\includegraphics[width=0.24\textwidth]{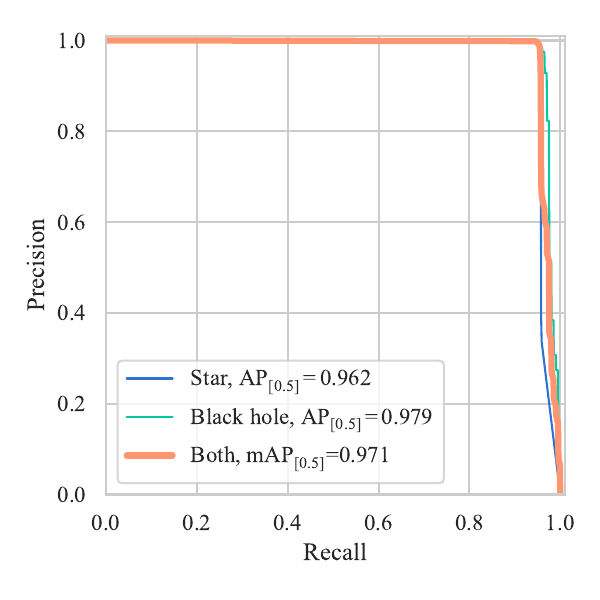}}
	\subfigure[Precision-Confidence Curve]{\includegraphics[width=0.24\textwidth]{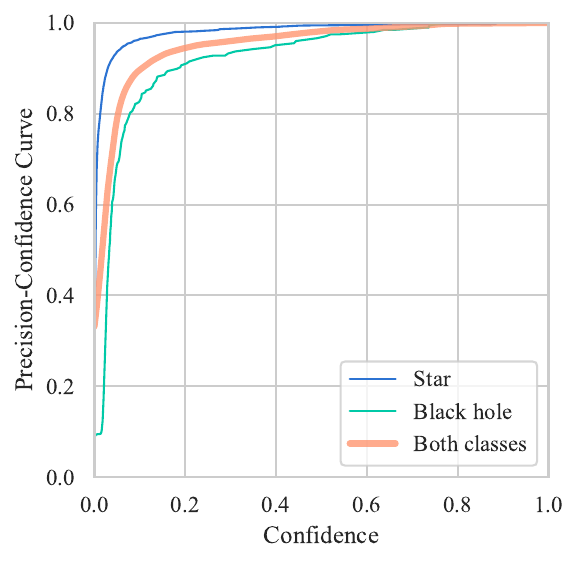}}
	\subfigure[Recall-Confidence Curve]{\includegraphics[width=0.24\textwidth]{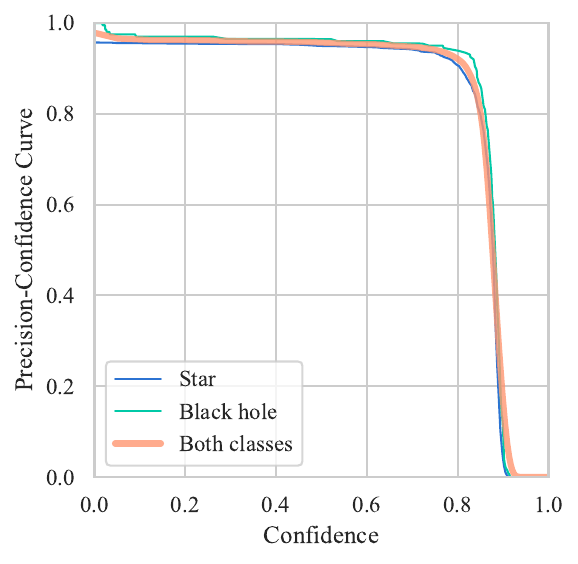}}
	\vspace{-0.4cm}
	\caption{F$_1$ scores{-confidence curve, presicion-recall curve, precision-confidence curve and recall-confidence curve} {}, where $\theta_\text{AD}=2.09\theta_c$ and the BH/star ratio is 1/10. {In subfigure (b), the Average Precision (AP) for black holes and stars with the IoU threshold of 0.5 are calculated by integrating from 0 to 1 and shown in the legend, respectively. And the mean Average Precision (mAP) is the average of AP for black holes and stars.}}
	\label{fig:f1_curve}
\end{figure*}

{

The working flows of our model are shown in Fig.~\ref{fig:flowchart}. Assume that our model outputs $N$ bounding circles, we will receive $N$ detected labels (BH or star) as well as their corresponding coordinates and confidence values.

Assume that the model's prediction is a black hole and its confidence value is $x$. Then we should also have a confidence level ranging from 0 to 1 to describe how cautious the prediction is. When $x < \text{confidence}$, the prediction is not valid and discarded. When $x > \text{confidence}$, the prediction is a black hole.
Then, we calculate the Intersection over Union (IoU) between the predicted circle and the ground truth circle. If the IoU is greater than a threshold (0.5 for example) and the label is correct, the prediction is considered correct.

Then all the $N$ detections from the model would be used to calculate the confusion matrix. The normalized confusion matrix is shown in Fig.~\ref{fig:confusion_matrix}.
There are actually three classes here: black hole, star, and background. Therefore, we have two sets of Precision, Recall, and F1 scores, which are all functions of confidence level and IoU threshold. When defining black holes as the positive class, stars and background are considered negative, yielding one set of precision, recall, and F$_1$ scores. When defining stars as the positive class, black holes and background are considered negative, yielding another set. The final precision, recall, and F$_1$ scores are the averages of these two sets.

As the confidence level increases, the model predicts more cautiously, and its predictions have higher credibility.
When we change the confidence level, the model's precision, recall and F$_1$ score will change, as shown in Fig.~\ref{fig:f1_curve}~(a)(c)(d). precision-recall curve is also shown in Fig.~\ref{fig:f1_curve}~(b), from which the average precision (AP) is calculated, which is the area under the curve. The mAP is the average of APs for black hole and star. The mAP$_{[0.5:0.95]}$ is the average of APs for all IoU thresholds from 0.5 to 0.95. The mAP$_{[0.5]}$ is the average of APs for IoU threshold 0.5. The mAP$_{[0.5:0.95]}$ is more comprehensive and accurate than mAP$_{[0.5]}$.
}

Since the effective variable affecting the resolution is the angular size of the accretion disk $\theta_\text{AD}$, we fix the observation distance and vary the size of the black hole accretion disk in practice, with the assumption that the accretion disk size is proportional to the black hole mass.
Four metrics are selected to measure the accuracy of the model, which are mAP$_{[0.5]}$ and mAP$_{[0.5:0.95]}$ for positioning capacity, and precision and recall for classification capacity. We have fixed the training period to 100 and the total images to 1000. For detailed configurations and hyperparameters of the model, see Table~\ref{tab:BHD_para} in the Appendix. The validation metrics with the change of training epoch are shown in the Appendix, where the angular size of the accretion disk is $1.78\theta_c$. It indicates that our model has a stable training process and a converged result.

\subsubsection{Parameter estimation model}\label{sec:reg}


To reduce computing time and power consumption, we utilized transfer learning for the convolutional layer in our model. Specifically, we used pre-trained weights from EfficientNet trained on ImageNet dataset for the convolutional layer in our regression and classification model. This approach resulted in improved accuracy values compared to using raw models with randomly initialized parameters. We chose the b1 model with 7.8 million parameters, which is practical for our experimental setup compared to the b5, b6, and b7 models with 30M, 43M, and 66M parameters, respectively.

The four fully connected layers are designed by ourselves, and the final output is the predicted parameter (e.g. $i$,$M$,$\phi$). Considering there are many ways to implement the model, the specific network architecture is shown in Fig.~\ref{fig:nn_structure}. The parameters of input and output are shown in Table~\ref{tab:model_reg} in the Appendix, where N is the batch size. Every fully connected layer follows a ReLU activation function and a dropout layer with a dropout rate of 0.5.

The loss function for $i, M$ is mean square error (MSE),
\begin{equation}
	L = \frac{1}{N}\sum_{n=1}^N (x_n-y_n)^2,
\end{equation}
where $N$ is number of objects and $x, y$ is prediction and ground truth respectively. For $\phi$ ranging from $[0,2\pi]$, the loss function is periodic MSE,
\begin{equation}
	L=\frac{1}{N} \sum_{n=1}^N \min\{\left(x_n-y_n\right)^2, \left(360\degree-x_n+y_n\right)^2\},
\end{equation}
and the metric for the regression task is mean absolute error (MAE)\footnote{We have also tested training with MAE as the loss function, but both the training speed and validation accuracy are not as good as MSE.}: $\quad l_n=\left|x_n-y_n\right|$ and $\text{MAE}= \operatorname{mean}(l_n)$. For the classification task, the loss function is cross-entropy loss,
\begin{equation}
	L=-\sum_{j=1}^4 y_j \log p_j,
\end{equation}
where $p_j$ is the predicted probability, $y_j$ is a bool value indicating whether the class label $j$ is the proper classification. In our work, there are four distinct temperatures of the accretion disk. And the metric for classification is accuracy.

The model is trained using 100 epochs and the 27~018 images. We have used Bayesian optimization to select the optimal hyperparameters, including learning rate, L2 regularization coefficients, and dropout rate during the training of the model. All subsequent results are from the models with optimal hyperparameters. The training system utilized a Gen Intel (R) i9-13900K with 24 vCPU cores and 128 GB of RAM, along with a single NVIDIA GeForce RTX 4070 with a 12 GB graphical processing unit. The environment includes Windows 11, Python 3.9.12, Torch 2.2.1, and other relevant software.

\section{TESTS}\label{sec:results}
\subsection{Unbalanced datasets}
In real observations, one of the challenges is that the datasets are unbalanced, where most of the objects are star and few are black holes. In these unbalanced datasets, conventional accuracy may be a misleading indicator, making our model evaluation a major challenge. Our solution is to make the black hole a positive class and set the proper confidence level to have a larger F$_1$ score. The F$_1$ scores of black holes, star and overall with the change of confidence are shown in Fig.~\ref{fig:f1_curve}. The F$_1$ score reaches the maximum of 0.97 when the confidence level is 0.625, which is close to the desired neutral 0.5. The F$_1$ score between 0.2 and 0.8 is flat, which indicates our model is insensitive to the change of confidence. These prove the good performance of our model in unbalanced datasets. So we simply choose the confidence level as 0.5 in the subsequent discussion.


To test the ability of our model to handle unbalanced datasets, we generate three groups of datasets, with the BH/star ratio of 1/3, 1/10 and 1/100 respectively, and $\theta_\text{AD}=1.6279\theta_c$. All other configurations are identical to the training process in section~\ref{sec:detection}. The results are shown in Table~\ref{tab:ratio}. When the ratio decreases, mAP also decreases because the unbalanced datasets cause unbalanced training. {

Since the final Precision and Recall are averages of those for black holes and stars, their values are influenced by both classes. When black holes are positive and the number of stars increases, FP rise, decreasing Precision. Conversely, when stars are positive and their number increases, FN rise, decreasing Recall.

The table shows that the final metrics primarily reflect the characteristics when stars are positive, indicated by increased Precision and decreased Recall. This is likely because the small number of black holes means changes in star numbers have little impact on Precision and Recall for black holes, but significantly affect those for stars.
}

{To sum up, even} if the dataset is unbalanced, the result remains satisfactory, indicating that our model is robust to unbalanced datasets.

\begin{table}[htbp]
	\centering
	\caption{Four metrics with the change of BH/star ratios, where $\theta_\text{AD}=1.6279\theta_c$.}
	\begin{tabular}{l@{\hskip 0.1in}c@{\hskip 0.1in}c@{\hskip 0.1in}c@{\hskip 0.1in}c}
		\hline \hline \addlinespace[2pt]
		BH/star & mAP$_{[0.5]}$ & mAP$_{[0.5:0.95]}$ & Precision & Recall  \\
		\addlinespace[1pt] \hline \addlinespace[2pt]
		1/3        & 0.97036       & 0.74807            & 0.91688   & 0.92440 \\
		1/10       & 0.95035       & 0.69731            & 0.95712   & 0.88908 \\
		1/100      & 0.90464       & 0.70239            & 0.95275   & 0.85548 \\
		\hline\hline
	\end{tabular}

	\label{tab:ratio}
\end{table}

\subsection{Angular size metrics}\label{subsec:angular_size}

It is important to analyze the influence of the resolution on the performance of the model. As a result, the model is trained under different $\theta$. We define the following regions: \textbf{ISCO range} denotes $\min{\theta_\text{ISCO}}<\theta_c<\max{\theta_\text{ISCO}}$, and \textbf{AD range} denotes $\min{\theta_\text{AD}}<\theta_c<\max{\theta_\text{AD}}$. They are all ranges rather than points because the masses of black holes in images are different. \textbf{Transition range} refers to the region in between.
\textbf{Normal resolution} (Super resolution) denotes that the black hole is larger (smaller) than $\theta_c$.
Since $r_{ISCO}$ < $r_{AD}$, it is clear that a larger angular size is needed to see a smaller object clearly. So the $\theta_{ISCO}$ range is larger than $\theta_{AD}$ range.

\begin{figure*}[htbp]
	\centering
	\includegraphics[width=0.49\textwidth]{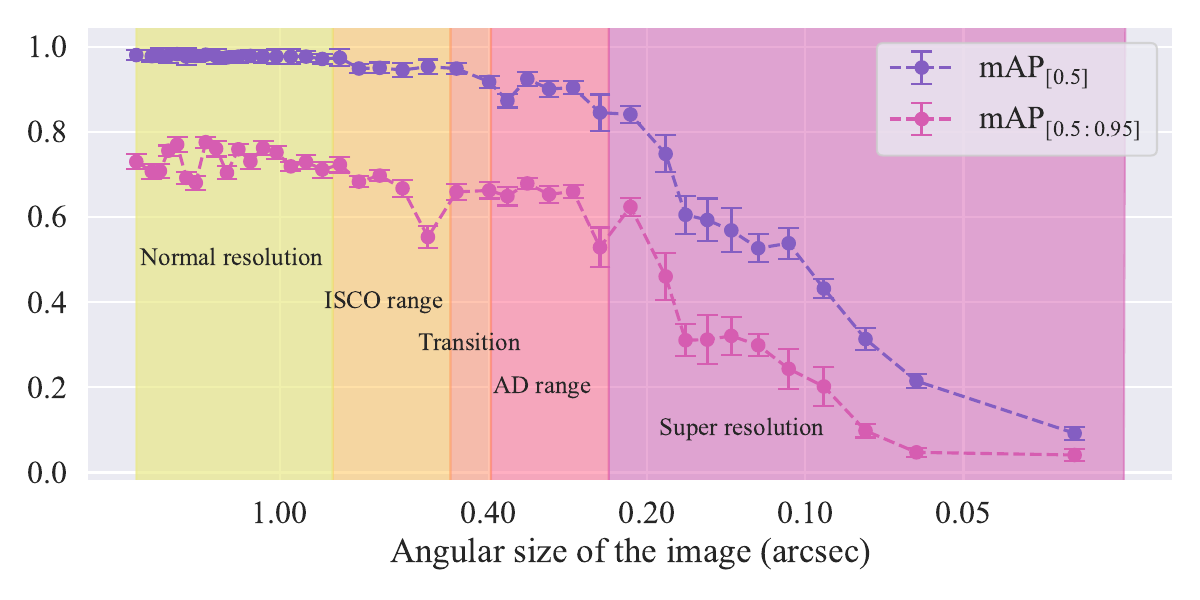}
	\includegraphics[width=0.49\textwidth]{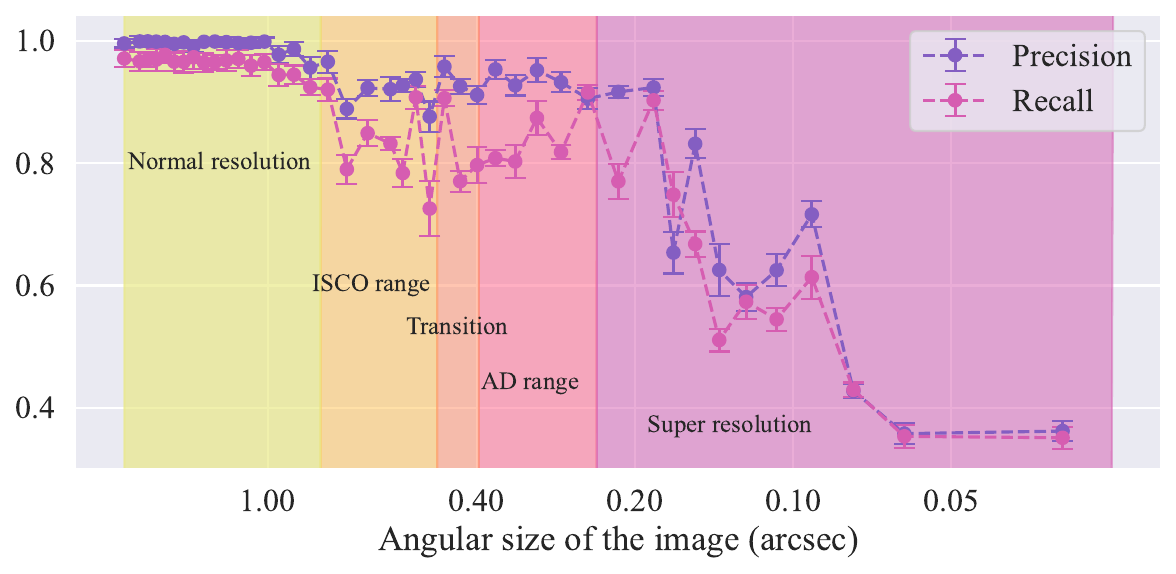}
	\vspace{-0.3cm}
	\caption{mAP$_{[0.5]}$, mAP$_{[0.5:0.95]}$, precision and recall for different $\theta$, where each color range denotes a region of iconic angular size, which is explained in section~\ref{subsec:angular_size}. {ISCO range: $[\min{\theta_\text{ISCO}}, \max{\theta_\text{ISCO}}]$. AD range: [$\min{\theta_\text{AD}}, \max{\theta_\text{AD}}]$. They are all ranges rather than points because the masses of black holes in images are different. Transition range is in between.
	Normal resolution (Super resolution) : black hole is larger (smaller) than $\theta_c$.}
	Note that, the x-axis is reversed to show the super-resolution region on the right.}
	\label{fig:mAP}
\end{figure*}
Considering the model has different metrics for different output parameters, we should have a unified metric defined in the range [0, 1].
For the detection model, the performance is defined as the mAP$_{[0.5]}$.
For the regression model, the performance is calculated in a normalized way: $1-\text{MAE} / \text{MAE}_{max}$, where $\text{MAE}_{max}$ is MAE of the {mean response. When the model has no informative training data, it defaults to predicting the mean of the target distribution, which can minimizes the mean absolute error (MAE), compared to predict other value instead. Mean response is the worst result we can get. For instance, for the inclination $i \in [-90\degree, 90\degree]$ with a uniform distribution. If the image has no information, the trained model would just guess $i=0$, the MAE is the maximum error, namely $45\degree$. Essentially, this is the maximum error we can get.}

For the classification of temperature, the performance is defined as the normalized accuracy: $\left(\text{Acc} - \text{Acc}_{min}\right) / \left(\text{Acc}_{max} - \text{Acc}_{min}\right)$, where $\text{Acc}_{min}$ and $\text{Acc}_{max}$ are the minimum and maximum accuracy, respectively. Accuracy is used here because our dataset is relatively balanced and the error are evenly distributed on both sides of the diagonal [cf. Fig.~\ref{fig:regression}].
If the model's performance is lower than the midpoint (mean of the max and the min), it is deemed to have lost its screening capability. 

To describe the requirement for the resolutions, we also define the \textbf{midpoint angle} as $\theta_{half}$ where the model has half of the performance, which is also the minimum resolvable angle. For example, $\theta_{half}$ for mAP is where $\text{mAP}=(\text{mAP}_\text{max}+\text{mAP}_\text{min}) / 2$. The results are shown in Table~\ref{tab:model_performance}. The first row is the model, the second column is the value at $\theta_{half}$. and the third column is the corresponding $\theta_{half}$.

\begin{table}
	\centering
	\caption{Models' half performances $\left(\text{in the range }[0, 1] \right)$ and corresponding minimum resolvable angle $\theta_{half}$.}
	\begin{tabular}{lcc}
		\hline\hline \addlinespace[3pt]
		Model                                                                      & Model's half performance & Corresponding $\theta_{half}$ \\
		\addlinespace[1pt] \hline \addlinespace[2pt]
		Detection\footnote{Calculated by mAP$_{[0.5]}$}                            & 0.596                    & $0.54\theta_c$                \\
		Regression\footnote{Calculated by the normalized MAE of inclination}       & 0.445                    & $1.48\theta_c$                \\
		Classification\footnote{Calculated by the model's accuracy of temperature} & 0.515                    & $0.69\theta_c$                \\
		\hline\hline
	\end{tabular}
	\label{tab:model_performance}
\end{table}

\begin{figure*}[htbp]
	\centering
	\includegraphics[width=0.99\textwidth]{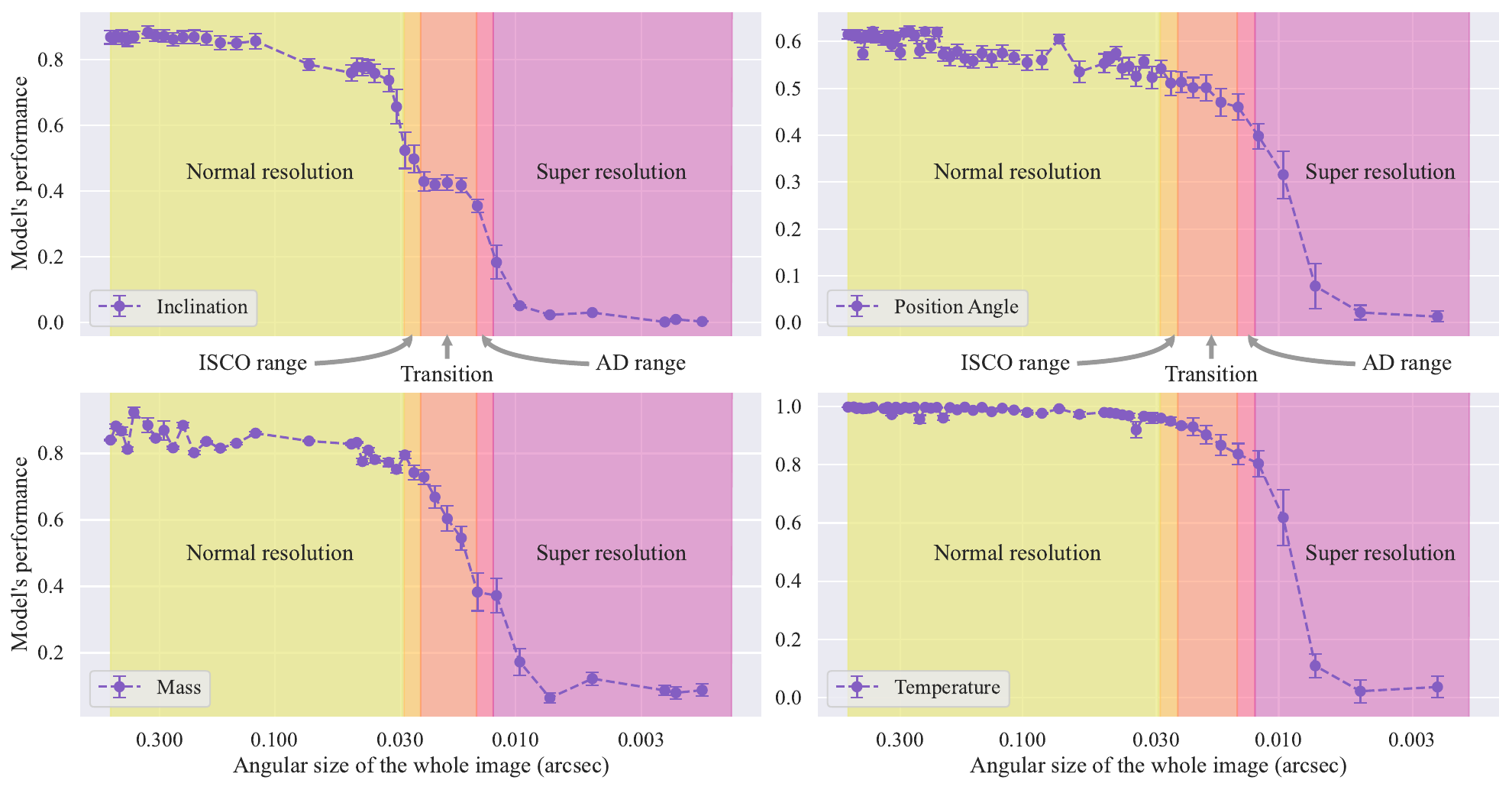}
	\vspace{-0.3cm}
	\caption{Performance of recognition model for different angular sizes of images, where each color range denotes a region of iconic angular size, which is explained in section~\ref{subsec:angular_size}. 
	{ISCO range: $[\min{\theta_\text{ISCO}}, \max{\theta_\text{ISCO}}]$. AD range: [$\min{\theta_\text{AD}}, \max{\theta_\text{AD}}]$. They are all ranges rather than points because the masses of black holes in images are different. Transition range is in between.
	Normal resolution (Super resolution) : Black hole is larger (smaller) than $\theta_c$.}
	The first row displays the model's normalized MAE for inclination and position angle. The second row displays the normalized MAE and accuracy for mass and temperature, respectively.
	Note that, the x-axis is reversed to show the super-resolution region on the right.}
	\label{fig:reg_test}
\end{figure*}

\begin{figure*}[htbp]
	\centering
	\includegraphics[width=0.99\textwidth]{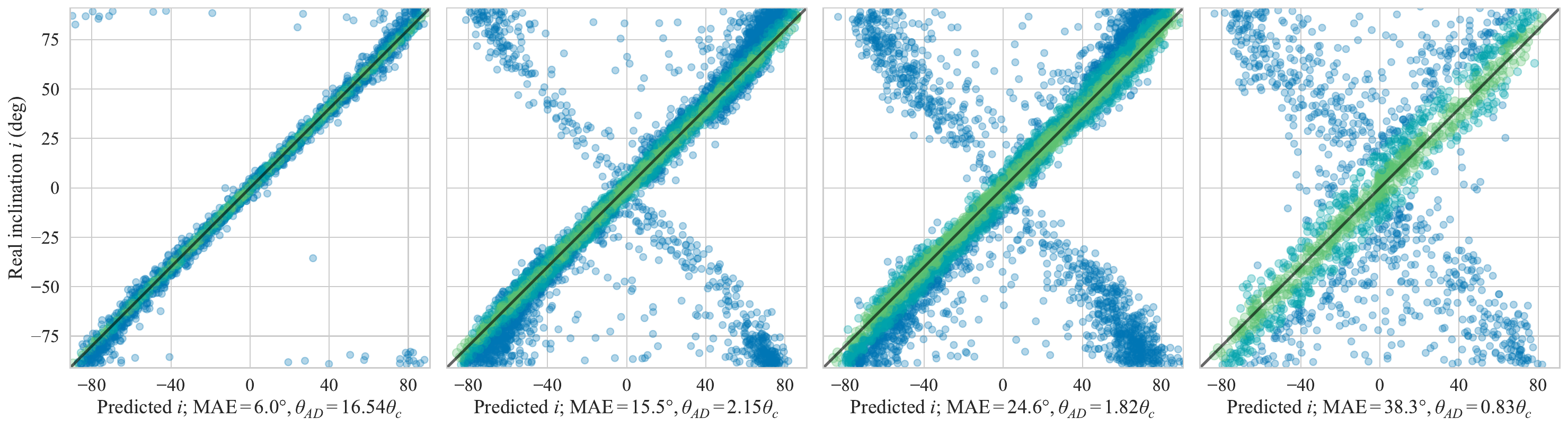}
	\includegraphics[width=0.99\textwidth]{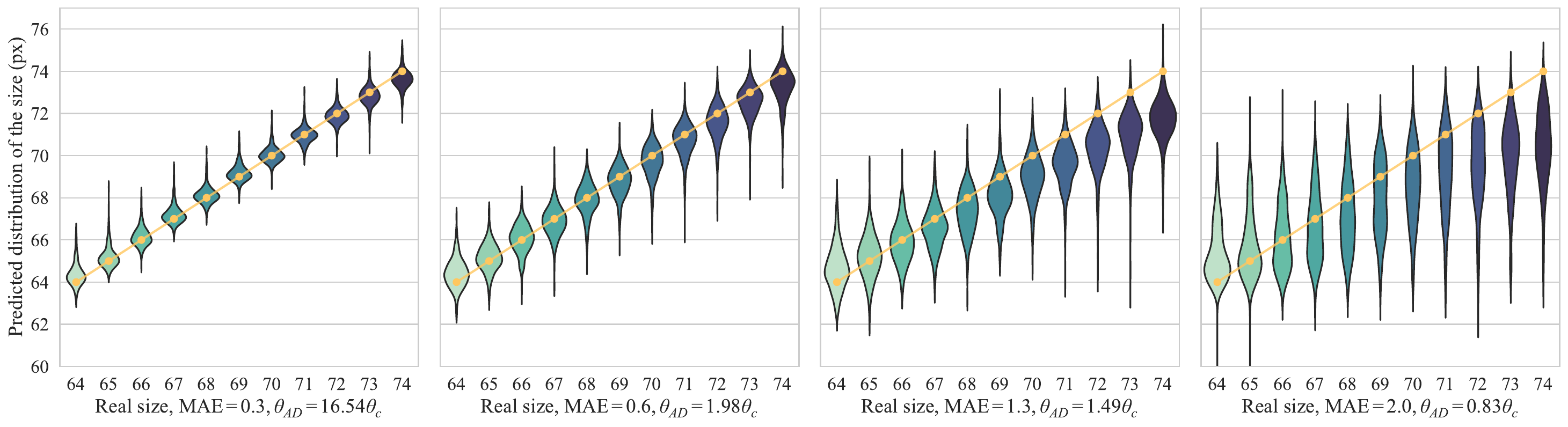}
	\includegraphics[width=0.99\textwidth]{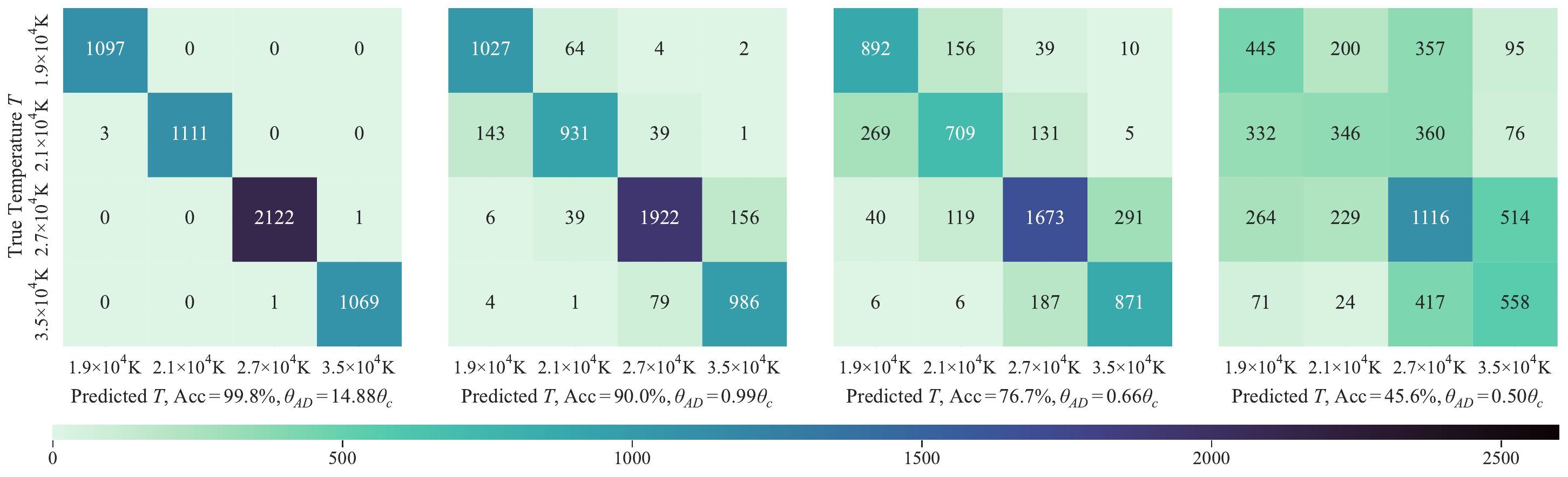}
	\caption{The ability of our model for parameter estimations. The first row shows the scatter plot of $i$, where the x-axis is the prediction and the y-axis is the ground truth. Green, blue-green and blue indicate errors belong to [0, $0.2$MAE], [$0.2$MAE, $0.5$MAE], $>0.5$MAE, respectively. The second row is a group of violin plots. The x-axis is the ground truth of size (px), which are integers ranging from 64 to 75. Each ``violin'' shows the distribution of prediction for each ground truth, and the yellow dots are the baselines. The third row shows the confusion matrix of the classification for $T$.}
	\label{fig:regression}
\end{figure*}
\begin{figure*}[htbp]
	\centering
	\includegraphics[width=0.99\textwidth]{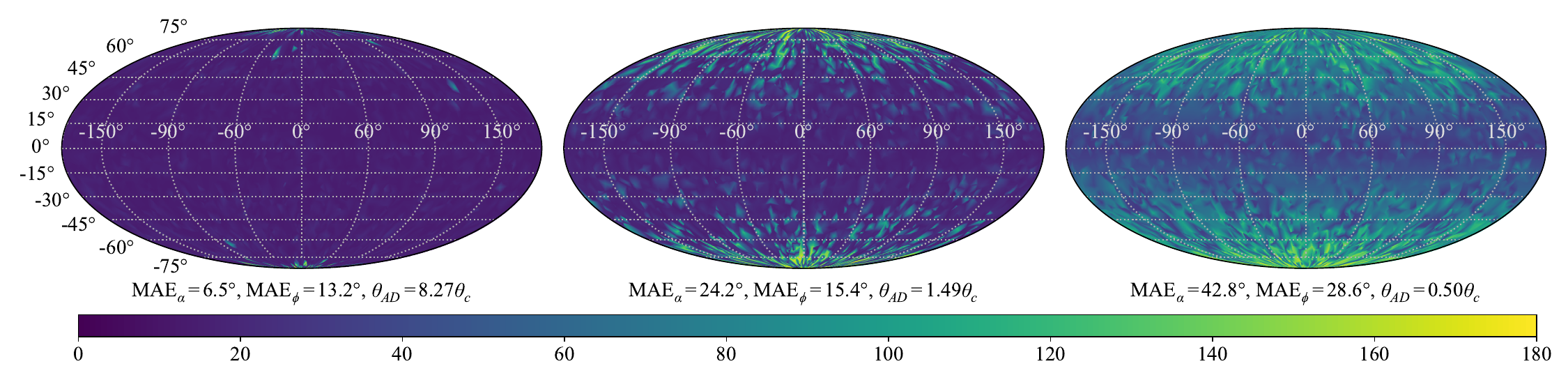}
	\caption{Skymap of absolute error of inclination and position angle, denoted by latitude and longitudes respectively. The color indicates the sum of absolute error for inclination and position angle (deg).}
	\label{fig:skymap}
\end{figure*}
Each metric with the change of $\theta$ for detection and recognition is shown in Fig.~\ref{fig:mAP} and Fig.~\ref{fig:reg_test}, respectively. For the detection model and classification model, the performance would retain a lot even when $\theta_\text{AD}=\theta_c$. For the BH detection, the model doesn't lose its ability until $\theta_\text{AD}=0.54\theta_c$ in terms of mAP$_{[0.5]}$. For the classification of temperature, the model still has the accuracy of 89\% when $\theta_\text{AD}=\theta_c$ and retains its functionality until $\theta_\text{AD}=0.69\theta_c$. The result shows that even if the shadow is indistinguishable in the context of the classical Rayleigh criterion, it can still be identified by our NN model, suggesting the properties of super-resolution detection of NNs \cite{SR}, which also indicates that our model has an exciting ability to extract every little information from the super-blurred image. However, for the estimation of $i$ and $M$, the model has not reached the edge of super-resolution. The model has half of its functionality when $\theta_\text{AD} \approx 1.5\theta_c$ (or $\theta_\text{ISCO} \approx \theta_c$). And it almost loses all of its ability when $\theta_\text{AD}$ reaches $\theta_c$. And for the estimation of $\phi$, the performance of our model is not that satisfactory. Although the model still has half of the functionality until $\theta_\text{AD} \approx 0.7\theta_c$, its overall performance is almost below 0.6. 
The probable reasons are as follows: The detection and estimation for $\phi$ and $T$ of black holes are mainly based on the outline shape and color scale of the image, so even if $\theta_\text{AD}<\theta_c$, some part of the information will still be retained.
For the regression of $i$ and $M$, the ability of our model starts to decline after the diffraction limit of the ISCO is reached ($\theta_\text{ISCO}<\theta_c$). When $\theta_\text{ISCO}<\theta_c$, the shadow will be connected to a facula and thus difficult to distinguish the inclination $i$. As for the estimation of $M$ (infer from the size of the shadow), when $\theta_\text{AD}>\theta_c$, the size of PSF is much larger than that of the shadow, so the size of the shadow in the image no longer depends on the size of the shadow itself but on the size of the PSF, which makes it difficult to estimate.





We have visualized the degree of conformity between prediction and ground truth for $i, M$ and $T$, see Fig.~\ref{fig:regression}. The first row is the scatter plot for $i$. The ``X'' shaped plots indicate that the MAE of $i$ goes up as $|i|$ increases. The second row is the violin plot for $M$ (inferred by the size of its shadow), which shows the distribution of prediction on the y-axis for each ground truth on the x-axis. The predictions gradually go diffuse and inaccurate as $\theta$ increases. The third row shows the confusion matrices for the classification of $T$. The data is distributed on the diagonal and spread out when $\theta$ increases. The error is shown as skymaps in Fig.~\ref{fig:skymap}, where latitude and longitudes denote $i$ and $\phi$ respectively. These plots show that errors are mainly distributed in the part with a larger inclination angle. The data of the skymap is obtained by piecewise linear interpolator for interpolation and nearest neighbor interpolator for extrapolation in scipy. The former is a method of triangulation of the input data using Qhull's method \cite{barber1996quickhull}, followed by the linear center of gravity interpolation on each triangle.

To sum up, our model achieves the high performance of black hole detection and parameter estimation by the maturity of a pre-trained YOLO, EfficientNet model and our proper modification.
According to the results above, minimum resolvable angular size and maximum observation distances obtained by different discriminants or models are shown in Table~\ref{tab:capability}, and observed distances correspond to a fixed black hole mass of $4\times10^4M_\odot$.
\begin{table}[htbp]
	\centering
	\caption{Min of $\theta_\text{AD}$ and max observable distance}
	\begin{tabular}{l@{\hskip 0.2in}l@{\hskip 0.2in}l}
		\hline \hline \addlinespace[2pt]
		Criterion                   & Resolution   & Max distance \\
		\hline \addlinespace[1pt]
		Rayleigh criterion          & 10.48$\mu$as & 83.08ly      \\
		Black hole detection        & 5.659$\mu$as & 153.9ly      \\
		Inclination estimation      & 15.51$\mu$as & 56.14ly      \\
		Mass estimation             & 15.93$\mu$as & 54.66ly      \\
		Position angle estimation & 7.126$\mu$as & 122.2ly     \\
		Temperature classification  & 7.231$\mu$as & 120.4ly      \\
		\hline \hline
	\end{tabular}
	\label{tab:capability}
\end{table}
Black holes that were ejected from the Hyades in the last 150 Myr display a median distance $\sim80$ pc (260.8 ly) from the Sun \cite{near}. Therefore, the methodology presented in this work might detect black holes in this range, according to Table~\ref{tab:capability}.

\subsection{Model tests with M87*}\label{sec:testing}
\begin{figure}[htbp]
	\centering
	\includegraphics[width=0.4\textwidth]{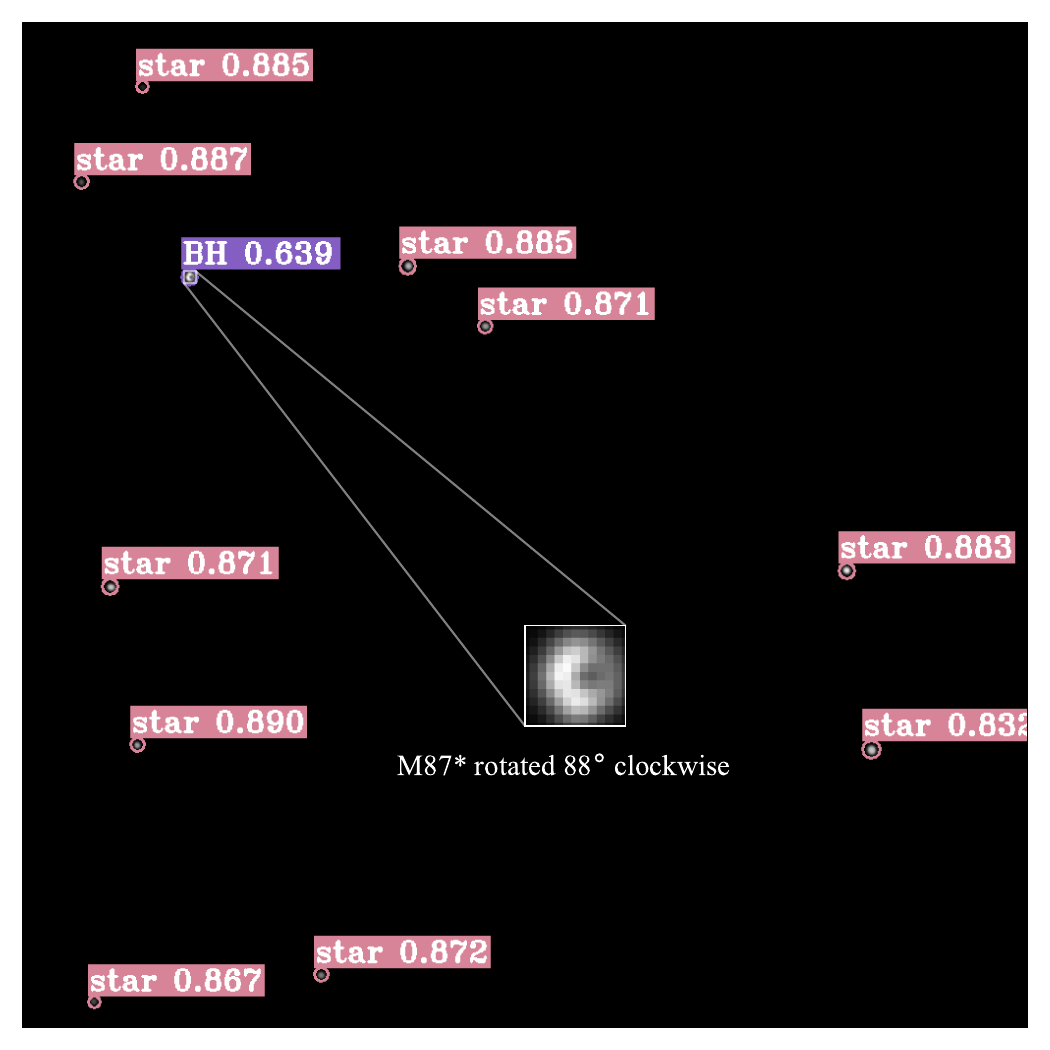}
	\caption{Model validation results, where the number after the label denotes the confidence of the prediction (between $[0,1]$), namely the probability that the object is a BH (star).}
	\label{fig:testing}
\end{figure}

Although the model performs well in simulated training, validation, and test sets, its real-world performance in detecting black hole shadows is what truly matters. To test the model's ability to detect real black holes, we scaled down an image of M87* observed by the EHT and added it to the generation pipeline along with other objects and background noise. The results are presented in Fig.~\ref{fig:testing}.

In this task, we first convert the M87* \cite{EHT1} black hole captured by the EHT into a grayscale image and compress it to $40\times40$, which is then fed into the data pipeline of the telescope simulation. We make the black hole's angular size 20$\mu as$ and rotate it clockwise by 88$\degree$ (the angle is randomly generated), accompanied by 10 star and random noise to ensure the SNR $<10$. The final image is input into the model which has been trained in the corresponding resolution in section~\ref{sec:model}, to get the output of the classification, location and confidence level. The result is shown in Fig.~\ref{fig:testing}, indicating that the model can successfully classify correctly all of a black hole and ten star, and accurately locate their positions. The confidence level of the black holes is 0.639, and for all the star is above 0.80, according to the output of the BH detection model.

\begin{figure}[htbp]
	\centering
	\includegraphics[width=0.43\textwidth]{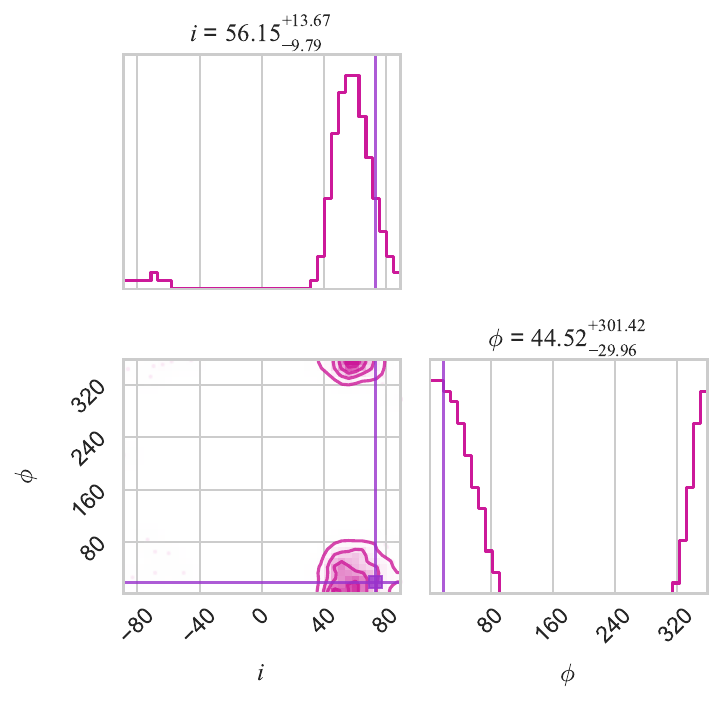}
	\vspace{-0.2cm}
	\caption{Parameter estimation result of M87*. The estimated value of $\phi$ is misleading because of the periodicity. The output of our model for $\phi$ is $31.9\degree$.}
	\label{fig:kde}
\end{figure}

The parameter estimation model is also tested. According to the EHT collaboration \cite{EHT5}, the position angle of M87* is $288\degree$ and the inclination angle is $17\degree$. In our coordinate system, take the transform $i \rightarrow (90\degree - i)$ and $\phi \rightarrow 90\degree - (360\degree - \phi)$ \footnote{The spin axis of the accretion disk in this work is vertical while in Ref.~\cite{EHT5} is horizontal. The positive rotation direction for $\phi$ in this work is counterclockwise while in Ref.~\cite{EHT5} is clockwise.}, and they should be $i_\text{true}=73\degree, \phi_\text{true}=18\degree$. The model outputs $i_\text{pred}=55.9\degree, \phi_\text{pred}=31.9\degree$. The posterior distribution of estimated parameters is shown in Fig.~\ref{fig:kde}. The posterior distribution is obtained by the distribution of ground truth from the test dataset that satisfies $i_\text{pred} \approxeq 55.9\degree$ and $\phi_\text{pred} \approxeq 31.9\degree$, where $\approxeq$ denotes the difference is less than $10\degree$. Our model performs better in terms of position angle but has a larger error for the estimation of the inclination.

\subsection{Model tests with real observation}

\begin{figure*}[htbp]
	\centering
	\includegraphics[width=0.98\textwidth]{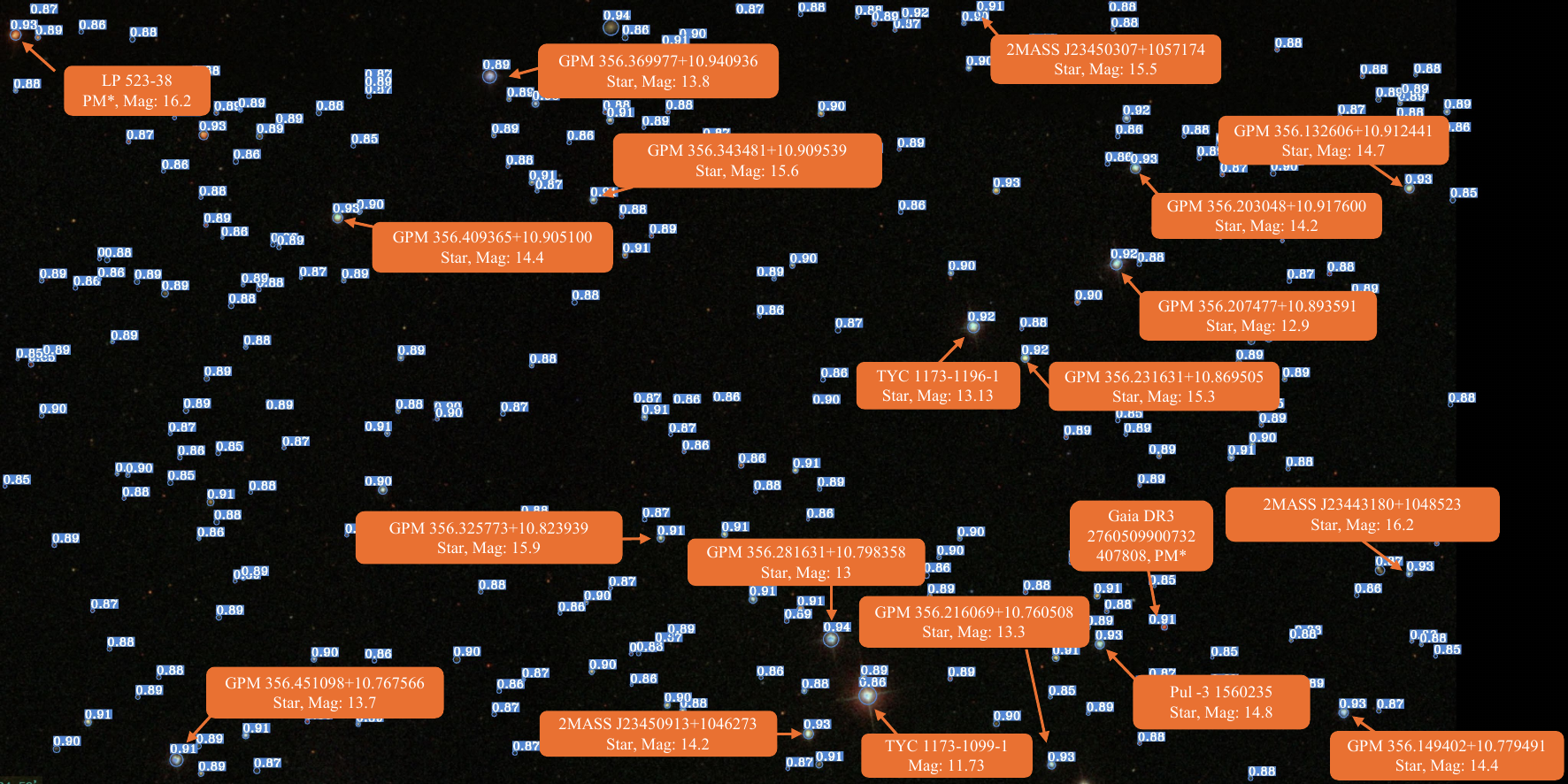}
	\caption{Star objects in the test dataset, where all the objects are labeled as stars by the model. The confidence value is shown in the blue box. All the verified stars is labeled with a orange box. PM* means High Proper Motion Star, which is a star that exhibits a significant change in its position on the sky over time due to its own motion through space relative to the Sun. For presentation purposes, the original colored image was used, but the images input into the model are in grayscale.}
	\label{fig:stellar_test}
\end{figure*}

To further validate our model, we selected observational data from the Hubble Space Telescope near the coordinates 23h44m56.761s+10d48m57.335s, with a field of view of $8.12 \times 4.61$ arcminutes, obtained from the SIMBAD database \cite{2000A&AS..143....9W}. After converting these images to grayscale, we applied our model for detection. Since there are no black holes in the image, all the model's predictions were classified as stars. At a 50\% confidence level, almost all luminous objects were labeled by the model, resulting in a cluttered image. Therefore, we chose an 85\% confidence level for display purposes, as shown in Fig.~\ref{fig:stellar_test}.

The figure demonstrates that the output labels of our model. However, only a few of the celestial bodies in the this observational image have been confirmed to be of specific types (stars, galaxies, quasars, etc.) according to previous works \cite{Yu_2022, Mager_2018, Skrutskie_2006, 2020yCat.1350....0G, 2011yCat.2306....0A}. The majority have not been verified. This makes it challenging to determine the accuracy of the predictions for the unverified objects. However, for verified stars, the model performed exceptionally well. It detects all the verified stars with relatively high confidence levels, most of which are above 90\%. The model's performance is consistent with the results of the test dataset, indicating that the model has a strong generalization ability and can accurately identify stars in observational data.

There are some discrepancies between simulated images and observational data, leading to certain prediction errors. For instance, some brighter stars exhibit diffraction spikes in observations, which the model can identify, but these affect the confidence level. In this image, the brightest star, TYC 1173-1099-1, has a predicted confidence level of 86\%, whereas some smaller stars have confidence levels up to 93\%. Additionally, the background noise in simulated images differs from real noise, which may also impact the model's performance. Despite these discrepancies, the model's performance is still satisfactory, indicating that our simulated images are realistic enough to applied to real-world tasks.

However, this result indicates that the difference between our simulated data and realistic scenarios is small enough that the model can still perform well in real-world situations.

\section{DISCUSSIONS AND CONCLUSIONS}\label{sec:discussion}

Our model is based on medium-sized, non-rotating black holes in the UV band while images of M87* taken by EHT \cite{EHT5} are based on the supermassive, rotating black hole in the radio band.
However, the difference in terms of spin and observation wavelength might not perform a significant role in the detection and parameter estimation task. Our NN model recognizes a black hole by its doughnut-like shape, which is nearly identical in the ratio band (see Fig.~1 in \cite{refId0}) and UV band (see Fig.~\ref{fig:diff2}). Additionally, according to the GRMHD simulation, the spin of the black hole mainly affects the size of its shadow rather than its shape at high temperatures (cf. the first row of Fig.~2 in \cite{EHT5}). That is the reason why our model can still get a decent result despite the huge difference between the model's training data and the real black hole. This indicates that the model has a certain degree of robustness and generalization ability.


Our model is underestimated by the calculations in section~\ref{sec:model}. Compressing a $3072\times3072$ image to $1024\times1024$ during image processing results in some loss of information in image quality for our model's input data. There is also a loss of color information when only considering the luminosity. Additionally, it is important to note that the actual black hole is a Kerr black hole, and the accretion disk of a Kerr black hole may be larger than that of a Schwarzschild black hole depending on the direction of rotation and other factors \cite{2012MNRAS.420..684P}. The size and temperature of a black hole's accretion disk are determined by various parameters, such as the accretion rate, which can vary depending on the environment surrounding the black hole \cite{Accetion}. This variability allows for the existence of larger black holes with larger accretion disks, which are easier to observe. The advancement of telescope manufacturing has led to the launch of larger and more advanced telescopes into space, such as the James Webb Space Telescope (JWST) \cite{Gardner_2006} with a 6.5m aperture. This development demonstrates that humans can launch larger optical telescopes with smaller imaging FWMH into space, expanding the observation range of the model. Additionally, the ensemble NN model is highly versatile. It can be used to detect black holes, as demonstrated in this paper, and can also be applied to other tasks, such as identifying other celestial objects or galaxies. One way to achieve this is by replacing the training data with simulation images of the objects. The model is applicable to other telescopes, including radio and optical interferometers operating in the ratio, infrared and visible wavelength bands. However, the telescope simulations presented in this paper should be replaced with simulation programs for the corresponding telescopes.

To sum up, this work presents an ensemble NN model with YOLO and EfficientNet as the backend. The model can detect and recognize black holes in both the simulated images and the real-world task, which has demonstrated that it can accurately work in real-world situations for detecting black holes and estimating parameters for potential candidates.

First, we have constructed a data pipeline consisting of accretion disk ray-tracing and telescope simulation. Realistically shaped black holes are obtained through reverse ray tracing. Telescopic simulations were then conducted, revealing that black holes are indistinguishable when their angular sizes of ISCO are smaller than the imaging FWHM. These simulated observations were ultimately used to train the ensemble NN model.

Using the dataset above, the model structure and loss function are altered based on the YOLO and EfficientNet as the backend, followed by training until convergence. For black hole detection, the model has a high detection performance, which achieves mAP$_{[0.5]}$ values of 0.9176 even when $\theta_\text{AD}$ reaching the imaging FWHM ($\theta_c$), and doesn't lose its detection ability until $0.54 \theta_c$, indicating that our detection model can go somewhat beyond the limits of the traditional Rayleigh diffraction limit. This is also the case for the estimation of $T$ and $\phi$, with the requirement of $\theta_\text{AD} \gtrsim 0.7\theta_c$.
In other words, super-resolution recognition beyond the traditional optical diffraction criterion is realized.
On the other hand, recognition for $i$ and $M$ requires a significantly higher resolution than detection, with a minimum requirement of $\theta_\text{AD} \gtrsim 1.5\theta_c$, which is natural, since estimating the parameters of black holes is more sophisticated than simply detection them and thus requires a higher resolution.

Our model was tested on observational data from both the Hubble Space Telescope and the Event Horizon Telescope (EHT). For the Hubble data near coordinates 23h44m56.761s +10d48m57.335s, the model successfully identified all stars with confidence levels mostly above 90\%.
Additionally, when tested on the image of M87* from the EHT, the model accurately distinguished the black hole with a confidence level of 0.639 and identified all stars with confidence levels above 0.8. These results demonstrate the model's strong generalization ability and its applicability across different observational data sets. However, there are some discrepancies between simulated images and observational data, which may affect the model's performance. For example, some brighter stars exhibit diffraction spikes in observations, which can impact the confidence level of the model's predictions. The background noise in simulated images also differs from real noise, which may affect the model's performance. For the test with M87*, the data is from radio band, which is different from our model.

Despite these discrepancies, the model's performance is still satisfactory, indicating that the difference between our simulated data and realistic scenarios is small enough that the model can still perform well in real-world situations.


In this paper, we do not consider other luminous objects such as galaxies and quasars, but they may interfere with the identification of black holes. For example, some galaxies might also show the shape of black holes, and larger galaxies may affect the imaging quality of the observation picture. To solve this issue, we can increase the complexity of the celestial body in the training data.
Additionally, interstellar dust may block high-energy ultraviolet rays, which can affect the accuracy of our observations.

In future work, it may be possible to obtain more realistic and accurate images of black holes by rendering Kerr black holes. The training data should include other celestial bodies such as galaxies and quasars to better simulate real-world observations. To reduce discrepancies between simulated images and observational data, we can use a portion of real observation data to the training data. The calculation of PSF should be refined to better model the diffraction effects, aberrations, and other imperfections of the telescope. The effect of stardust should also be considered. Additionally, Bayesian statistics can be used to compute the posterior distribution of the parameters in parameter estimation, instead of only computing the parameter values.

\section{ACKNOWLEDGMENTS}

The authors gratefully acknowledge Shangyu Wen for insightful talks and input throughout the study.
This work is supported by the National Natural Science Foundation of China (NSFC) with Grant Nos. 12175212, 12275183 and 12275184.
The simulation of black holes references GitHub repo with code available at \cite{Charles2013}. The conclusions and analyses presented in this publication were produced using the following software: Python (Guido van Rossum 1986) \cite{10.5555/1593511}, Opencv (Intel, Willow Garage, Itseez) \cite{itseez2015opencv}, Scipy (Jones et al. 2001) \cite{2020SciPy-NMeth}, PyTorch  (Meta AI September 2016) \cite{NEURIPS2019_9015}, Matplotlib (Hunter 2007) \cite{Hunter:2007}, Seaborn \cite{Waskom2021} and Corner (Daniel et al. 2016) \cite{corner}. {This research has also made use of the SIMBAD database \cite{2000A&AS..143....9W}, CDS, Strasbourg Astronomical Observatory, France.}
This work is finished on the server from Kun-Lun in Center for Theoretical Physics, School of Physics, Sichuan University.

\appendix
\setcounter{secnumdepth}{0}

\section{APPENDIX}\label{appendix:model}
Each output image in section~\ref{sec:simulation} is of size $3072\times3072$, exactly the pixel number of CCD. It is subsequently compressed to size $1024\times1024$. The reason for not using the $1024\times1024$ image directly is that due to the short UV wavelength, the continuum spectrum of the PSF will show a very sharp peak, and if the input image is small, the sampling interval will be too large during sampling, resulting in sampling distortion. After testing, the size of $3072\times3072$ is just enough to meet the requirements, see Fig.~\ref{fig:telescope} (b).

\begin{table}[htbp]
	\centering
	\caption{Example of a labeled file}
	\begin{tabular}{c@{\hskip 0.1in}@{\hskip 0.1in}c@{\hskip 0.1in}c@{\hskip 0.1in}c}
		\hline \hline \addlinespace[1pt]
		Class & $x$ coord. & $y$ coord. & radius   \\
		\hline \addlinespace[1pt]
		1     & 0.431429   & 0.8350     & 0.015714 \\
		0     & 0.240357   & 0.5335     & 0.016429 \\
		0     & 0.761071   & 0.6615     & 0.016429 \\
		0     & 0.037500   & 0.5605     & 0.010714 \\
		0     & 0.325000   & 0.5580     & 0.010000 \\
		0     & 0.594643   & 0.0225     & 0.009286 \\
		\hline \hline
	\end{tabular}
	\label{tab:labels}
\end{table}

The example of labels for the detection model is shown in Table~\ref{tab:labels}, where the first line of the table indicates that there is a bounding circle for the black hole in the coordinate (0.43, 0.83) and radius 0.016, each value relative to the whole image.
(``1" accounts for black holes and ``0" accounts for star.)


Since we have changed the original bounding boxes of the YOLO model to bounding circles, recalculation of IoU is needed. First, find the distance between the centers of two circles $d$.
Check for three conditions:
If $d>r_1+r_2$, the circles do not intersect.
If $d \leq|r_1-r_2|$, one circle is completely inside the other.
Otherwise, the circles intersect, and you need to calculate the area of intersection.
The area of intersection $\left(A_{\text {intersection }}\right)$ :
\begin{equation}
	\begin{aligned}
		A_{\text {intersection }}= & r_1^2  \arccos\left(\frac{d^2+r_1^2-r_2^2}{2  d  r_1}\right)+ \\		                        & r_2^2  \arccos\left(\frac{d^2+r_2^2-r_1^2}{2  d  r_2} \right)- \\
		\frac{1}{2}                & ((-d+r_1+r_2)  (d+r_1-r_2)                                    \\
		                           & (d-r_1+r_2) (d+r_1+r_2))^{1/2},
	\end{aligned}
	\label{eq:intersection}
\end{equation}


The area of the union $\left(A_{\text {union }}\right)$ is $A_{\text {union }}=\pi  r_1^2+\pi  r_2^2-A_{\text {intersection }}$. Finally, calculate the loU:
\begin{equation}
	\mathrm{IoU}=\frac{A_{\text {intersection }}}{A_{\text {union }}}
	\label{eq:iou}
\end{equation}
\begin{table}
	\centering
	\caption{Model structure of recognition model}
	\begin{tabular}{c@{\hskip 0.2in}c@{\hskip 0.2in}c}
		\hline \hline \addlinespace[2pt]
		Name     & Type              & Output size       \\
		\addlinespace[1pt] \hline \addlinespace[2pt]
		Initial  & Input image       & $(N,240,240,3)$   \\
		Eff. Top & Pre-trained model & $(N, 8, 8, 1280)$ \\
		Avg pool & Global Avg. Pool  & $(N, 1280)$       \\
		Dropout  & Dropout layer     & $(N, 1280)$       \\
		FC1      & Linear+ReLU       & $(N, 256)$        \\
		Dropout  & Dropout layer     & $(N, 256)$        \\
		FC2      & Linear+ReLU       & $(N, 32)$         \\
		Dropout  & Dropout layer     & $(N, 32)$         \\
		FC3      & Linear+ReLU       & $(N, 1)$          \\
		\hline \hline
	\end{tabular}
	\label{tab:model_reg}
\end{table}
\begin{table}[htbp]
	\centering
	\caption{Parameterization of the BH detector model}
	\begin{tabular}{cc|cc}
		\hline \hline
		Hyper-Para    & Value  & Hyper-Para   & Value \\
		\hline
		Learning Rate & 0.01   & epochs       & 100   \\
		Momentum      & 0.937  & image size   & 1024  \\
		Weight Decay  & 0.0005 & Augmentation & True  \\
		Batch Size    & 16     & Pre-Trained  & True  \\
		\hline\hline
	\end{tabular}

	\label{tab:BHD_para}
\end{table}

The BH detector model is obtained after tuning hyperparameters. The training is started with the pre-trained weights using the ImageNet Train dataset \cite{5206848} provided by the Ultralytics YOLOv5 project. The optimizer is set to the stochastic gradient descent (SGD) and the optimal parameters are shown in Table~\ref{tab:BHD_para}:



\nocite{*}
\bibliography{shadow}

\end{document}